\documentclass{article}
\usepackage{epsfig}
\usepackage{rotating}

\newcommand{\code}[1]{{\tt #1}}

\newcommand{\ft}{\tilde{f}}

\newcommand{\snu}{\tilde{\nu}}

\newcommand{\gt}{\tilde{g}}

\newcommand{\into}{\rightarrow}
        
\newcommand{\nn}{\nonumber}
\newcommand{\beq}{\begin{equation}}
\newcommand{\eeq}{\end{equation}}
\newcommand{\bea}{\begin{eqnarray}}
\newcommand{\eea}{\end{eqnarray}}

\newcommand{\beqa}{\begin{eqnarray}}
\newcommand{\eeqa}{\end{eqnarray}}

\newcommand{\app}[3]{Astropart.\ Phys.\ {\bf #1} (#2) #3}

\newcommand{\plb}[3]{Phys.\ Lett.\ {\bf B#1} (#2) #3}
\newcommand{\npb}[3]{Nucl.\ Phys.\ {\bf B#1} (#2) #3}
\newcommand{\ibid}[3]{{\em ibid.}\ {\bf B#1} (#2) #3}
\newcommand{\cpc}[3]{Comm.\ Phys.\ Comm.\ {\bf #1} (#2) #3}
\newcommand{\prl}[3]{Phys.\ Rev.\ Lett. {\bf #1} (#2) #3}
\newcommand{\apj}[3]{Astrophys.\ J.\ {\bf #1} (#2) #3}
\newcommand{\prd}[3]{Phys.\ Rev.\ {\bf D#1} (#2) #3}

\newcommand{\href}[2]{#1}

\def\rn{\noindent\parshape 2 0truecm 8.5truecm 0.3truecm 8.2truecm}
\def\rn{}
\def\nn#1 #2{#2. #1}                            
\def\nnn#1 #2 #3{#2. #3. #1}                    
\def\nnnn#1 #2 #3 #4{#2. #3. #4. #1}            
\def\nnnnn#1 #2 #3 #4 #5{#2. #3. #4. #5. #1}    

\def\rf#1;#2;#3;#4;#5 {{\frenchspacing\par\rn#1, #3 {\bf #4}, #5 (#2). \par}}
\def\rfbook#1;#2;#3;#4;#5 {{\frenchspacing\par\rn#1, {\it #3} (#5, 
#4, #2).\par}}
\def\rfprep#1;#2;#3 {{\frenchspacing\rn#1, #3 (#2);\ }}
\def\rfprepend#1;#2;#3 {{\frenchspacing\rn#1, #3 (#2).}}

\def\lsim{\mathrel{\rlap{\lower4pt\hbox{\hskip1pt$\sim$}}
     \raise1pt\hbox{$<$}}}         
\def\gsim{\mathrel{\rlap{\lower4pt\hbox{\hskip1pt$\sim$}}
     \raise1pt\hbox{$>$}}}         
\def\esim{\mathrel{\rlap{\raise2pt\hbox{$\sim$}}
     \lower1pt\hbox{$-$}}}         

\newcommand{\ds}{{\sf DarkSUSY}}
\newcommand{\dsver}{4.1}

\begin{document}
\null
\vskip .3cm
\begin{center}
{\bf \large DarkSUSY: Computing Supersymmetric Dark Matter Properties\\ 
Numerically}
\end{center}
\vskip .3cm
\centerline{P. Gondolo\footnote{paolo@physics.utah.edu}}
\centerline{\em Department of Physics, University of Utah, }
\centerline{\em 115 South 1400 East, Suite 201, Salt Lake City, UT 84112-0830, USA}
\centerline{J. Edsj\"o\footnote{Supported by the Swedish Research
Council (VR), edsjo@physto.se}}
\centerline{\em Department of Physics, Stockholm University,}
\centerline{\em AlbaNova University Center, SE-106~91~Stockholm, Sweden}
\centerline{P. Ullio\footnote{ullio@sissa.it}}
\centerline{\em SISSA, via Beirut 4, 34014 Trieste, Italy}
\centerline{L. Bergstr\"om\footnote{Partially supported by the Swedish Research
Council (VR), lbe@physto.se}}
\centerline{\em Department of Physics, Stockholm University,}
\centerline{\em AlbaNova University Center,  SE-106~91~Stockholm, Sweden}
\centerline{M. Schelke\footnote{schelke@physto.se}}
\centerline{\em Department of Physics, Stockholm University,}
\centerline{\em AlbaNova University Center, SE-106~91~Stockholm, Sweden}
\centerline{E.A. Baltz\footnote{eabaltz@slac.stanford.edu}}
\centerline{\em KIPAC,
Stanford University}
\centerline{\em P.O. Box 90450,MS 29
Stanford, CA 94309, USA}

\bigskip

\begin{abstract}
The question of the nature of the dark matter in the Universe remains
one of the most outstanding unsolved problems in basic science. One of the
best motivated particle physics candidates is the lightest supersymmetric
particle, assumed to be the lightest neutralino - a linear 
combination of the supersymmetric
partners of the photon, the Z boson and neutral scalar Higgs particles.
Here we describe \ds, a publicly-available advanced numerical package for
neutralino dark matter calculations.  In \ds\ one can compute the neutralino density in the Universe today  using precision methods which include resonances,
pair production thresholds and  coannihilations. Masses and mixings of 
supersymmetric particles can be computed within \ds\ or with the help of external programs such as \code{FeynHiggs}, \code{ISASUGRA} and \code{SUSPECT}. 
Accelerator
bounds can be checked to identify viable dark matter candidates. \ds\ also computes a large variety of astrophysical signals from neutralino dark matter, such as direct detection in low-background counting experiments and indirect detection
through antiprotons, antideuterons, gamma-rays and positrons from the Galactic halo or high-energy
neutrinos from the center of the Earth or of the Sun.
Here we describe the physics behind the package. A detailed manual will be provided with the computer package.
\end{abstract}

\noindent {\small Keywords:
{\em Supersymmetry; Dark Matter; Cosmology; Neutrino Telescopes; Gamma-ray
Telescopes; Cosmic Antiprotons; Cosmic Antideuterons; Cosmic Positrons; Direct Detection;
Indirect Detection; Numerical Code; Galactic Halo}}

\newpage
\tableofcontents
\newpage


\section{Introduction}
\label{sec:intro}
During the past few years, remarkable progress has been made in
cosmology, both observationally and theoretically.  One of the
outcomes of these rapid developments is the increased confidence that
most of the mass of the observable Universe is of an unusual form,
i.e., not made up of ordinary baryonic matter.  Recent
analyses combining high-redshift supernova luminosity distances,
microwave background fluctuations and the dynamics and baryon fraction
of galaxy clusters indicate that the present mass density of matter in
the Universe $\Omega_M=\rho_M/\rho_{\rm crit}$ normalized to the critical
density $\rho_{\rm crit}=3H_0^2/(8\pi G_N)=h^2\times 1.9\cdot 10^{-29} \ {\rm
g\, cm}^{-3}$ is $0.1 \lsim \Omega_Mh^2 \lsim 0.2$, which is
considerable higher than the value $\Omega_Bh^2\lsim 0.023$ allowed by
big bang nucleosynthesis \cite{tytler}.  Here $h\simeq
0.7\pm 0.15$ is the present value of the Hubble constant in units of
$100$ km s$^{-1}$ Mpc$^{-1}$.  A recent addition to the wealth
of experimental data which support the existence of non-baryonic
dark matter is the WMAP microwave background results \cite{wmap1}.
In a joint analysis of the WMAP data together with 
other CMBR experiments, large-scale structure data, supernova data and
the HST Key Project, the WMAP team report \cite{wmap1} 
$\Omega_Bh^2=0.0224\pm 0.009$,
$\Omega_Mh^2=0.135\pm 0.009$, and $\Omega_\Lambda = 0.73\pm 0.04$. Subtracting the baryonic contribution $\Omega_B h^2$ from the matter density $\Omega_M h^2$ leaves a non-baryonic cold dark matter density $\Omega_{CDM} h^2 = 0.113 \pm 0.009$.

Also from the point of view of structure formation, non-baryonic dark
matter seems to be necessary, and the main part of it should consist
of particles that were non-relativistic at the time when structure
formed (cold dark matter, CDM), thus excluding light neutrinos.  
Under reasonable assumptions,
the WMAP collaboration, using also galaxy survey and Ly-$\alpha$ forest data,
  limit the contribution of neutrinos to $\Omega_\nu h^2 < 0.0076
$ (95 \% c.l.). 

A well-motivated particle physics
candidate which has the required properties is the lightest
supersymmetric particle, assumed to be a neutralino
\cite{goldberg,krauss,ellis}.  (For thorough reviews of
supersymmetric dark matter, see \cite{jkg,lbreview}.)  Although supersymmetry
is generally accepted as a very promising enlargement of the Standard
Model of particle physics (for instance it would solve the so-called hierarchy problem
which essentially consists of understanding why the electroweak scale
is protected against Planck-scale corrections), little is known about
what a realistic supersymmetric model would look like in its details.
Therefore, it is a general practice to use the simplest possible
model, the minimal supersymmetric enlargement of the Standard Model
(the MSSM), usually with some additional simplifying assumptions.  Of
course, there is no compelling reason why the actual model, if nature
is supersymmetric at all, should be of this simplest kind.  However,
the MSSM serves as a useful template with which to test current ideas
about detection, both in particle physics accelerators and 
in dark matter experiments
and contains many features which are expected to be universal for any
supersymmetric dark matter model.  In fact, the knowledge gained by
studying the MSSM in detail may be of even more general
use, since it provides one specific example of a WIMP, a weakly
interacting massive particle, which is generically what successful
particle dark matter models require.\footnote{A completely different
type of particle is the axion \cite{raffelt}, which is very light, but
was never in thermal equilibrium.  Its phenomenology is very different
and will not be treated here.}

Over several years, we have developed analytical and numerical tools
for dealing with the sometimes quite complex calculations necessary
to go from given input parameters in the MSSM to actual quantitative
predictions of the relic density of the neutralinos in the Universe, and
the direct and indirect detection rates. The program package,
which we have named \ds, has now reached a high level of sophistication
and maturity, and we have released it publicly for the benefit of the 
scientific community working with problems related to dark matter.
This paper describes the basic structure and the underlying physical
and astrophysical formulas contained in \ds, as well as examples of its use.
The version of the package described in this paper is \ds\ \dsver.

For download of the latest version of \ds\ and for a more technical
manual, please visit the official \ds\ website, {\tt
http://www.physto.se/\~{}edsjo/darksusy/}.

\section{Definition of the Supersymmetric model}
\label{sec:MSSMdef}

We work in the framework of the minimal supersymmetric extension of
the Standard Model defined by, besides the particle content and gauge
couplings required by supersymmetry, the superpotential (the notation
used is that of \cite{bg} which marked the beginning of the
development of \ds, and is similar to~\cite{haberkane})
\begin{eqnarray}
   W = \epsilon_{ij} \left(
   - {\bf \hat{e}}_{R}^{*} {\bf Y}_E {\bf \hat{l}}^i_{L} {\hat H}^j_1
   - {\bf \hat{d}}_{R}^{*} {\bf Y}_D {\bf \hat{q}}^i_{L} {\hat H}^j_1
   + {\bf \hat{u}}_{R}^{*} {\bf Y}_U {\bf \hat{q}}^i_{L} {\hat H}^j_2
   - \mu {\hat H}^i_1 {\hat H}^j_2
   \right)
   \label{superpotential}
\end{eqnarray}
and the soft supersymmetry-breaking potential
\begin{eqnarray}
   \label{Vsoft}
   V_{{\rm soft}} & = &
   \epsilon_{ij} \left(
    -{\bf \tilde{e}}_{R}^{*} {\bf A}_E {\bf Y}_E {\bf \tilde{l}}^i_{L}
H^j_1
   - {\bf \tilde{d}}_{R}^{*} {\bf A}_D {\bf Y}_D {\bf \tilde{q}}^i_{L}
H^j_1
   + {\bf \tilde{u}}_{R}^{*} {\bf A}_U {\bf Y}_U {\bf \tilde{q}}^i_{L}
H^j_2
   - B \mu H^i_1 H^j_2 + {\rm h.c.}
   \right) \nonumber \\ &&
   + H^{i*}_1 m_1^2 H^i_1 + H^{i*}_2 m_2^2 H^i_2
   \nonumber \\ && +
   {\bf \tilde{q}}_{L}^{i*} {\bf M}_{Q}^{2} {\bf \tilde{q}}^i_{L} +
   {\bf \tilde{l}}_{L}^{i*} {\bf M}_{L}^{2} {\bf \tilde{l}}^i_{L} +
   {\bf \tilde{u}}_{R}^{*} {\bf M}_{U}^{2} {\bf \tilde{u}}_{R} +
   {\bf \tilde{d}}_{R}^{*} {\bf M}_{D}^{2} {\bf \tilde{d}}_{R} +
   {\bf \tilde{e}}_{R}^{*} {\bf M}_{E}^{2} {\bf \tilde{e}}_{R}
   \nonumber \\ && +
   {1\over2} M_1 \tilde{B} \tilde{B} +
   {1\over2} M_2 \left( \tilde{W}^3 \tilde{W}^3 +
         2 \tilde{W}^+ \tilde{W}^- \right) +
   {1\over2} M_3 \tilde{g} \tilde{g} .
\end{eqnarray}
We give these and the following expressions since they contain our sign 
conventions. It should be noted that various authors use various sign
conventions, and many errors, often difficult to find, can be avoided 
by keeping careful track of the signs, as we have tried to do 
consistently in \ds. 
Here $i$ and $j$ are SU(2) indices ($\epsilon_{12} = +1$). The Yukawa
couplings ${\bf Y}$, the soft trilinear couplings ${\bf A}$ and the
soft sfermion masses ${\bf M}$ are $3\times3$ matrices in generation
space. $\bf \hat{e}$, $\bf \hat{l}$, $\bf \hat{u}$, $\bf \hat{d}$ and
$\bf \hat{q}$ are the superfields of the leptons and sleptons and of
the quarks and squarks. A tilde indicates their respective scalar
components.  The $L$ and $R$ subscripts on the sfermion fields refer
to the chirality of their fermionic superpartners. $\tilde{B}$,
$\tilde{W}^3$ and $\tilde{W}^\pm$ are the fermionic superpartners of
the U(1) and SU(2) gauge fields and $\tilde{g}$ is the gluino field. $\mu$ is
the Higgsino mass parameter, $M_1$, $M_2$ and $M_3$ are the gaugino
mass parameters, $B$ is a soft bilinear coupling, while  $m^2_{1}$ and $m^2_{2}$ are
Higgs mass parameters.

These input parameters  are contained in common
blocks in the program.  The full set of input parameters in version 4.1 of \ds, to be given at the weak scale, is
$m_A$, $\tan\beta$, $\mu$, $M_1$, $M_2$, $M_3$, $A_{Eaa}$,
$A_{Uaa}$, $A_{Daa}$, $M^2_{Qaa}$, $M^2_{Laa}$, $M^2_{Uaa}$,
$M^2_{Daa}$, $M^2_{Eaa}$ (with $a=1,2,3$).  
The user may either provide these parameters directly to \ds\, or take advantage
of the implementation of a MSSM pre-defined through a reduced number of
parameters. In this model,
the basic set of parameters is $\mu$, $M_2$, $m_A$, $\tan\beta$, $m_0$,
$A_t$, $A_b$. Here $m_A$ is the mass of the CP-odd Higgs boson and $\tan\beta$ denotes the ratio, $v_2/v_1$, of the vacuum expectation values of the two neutral components of the SU(2) Higgs doublets. The parameters $m_0$, $A_t$ and $A_b$ are defined through the simplifying Ansatz: ${\bf M}_Q = {\bf M}_U = {\bf M}_D = {\bf M}_E = {\bf M}_L = m_0{\bf 1}$, ${\bf A}_U = {\rm diag}(0,0,A_t)$, ${\bf A}_D = {\rm diag}(0,0,A_b)$, ${\bf A}_E = {\bf 0}$.

Below we will give some details to clarify our convention and additional features.
Relevant quantities for phenomenological studies, such as the particle masses and
mixings, are consistently computed by \ds\ and available in arrays. The
supersymmetry part of the program can thus be used for many applications,
in particular for accelerator-based physics studies.
Particle decay widths are also
available, but currently only the widths of the Higgs bosons
are calculated, the other particles having fictitious widths of 1 or 5 GeV (for
the sole purpose of regularizing annihilation amplitudes close to poles). 

\subsection{Neutralino and chargino sectors}
The neutralinos $\tilde{\chi}^0_i$ are linear combinations of the
superpartners of the neutral gauge bosons, ${\tilde B}$, $\tilde{W}_3$
and of the neutral Higgsinos, ${\tilde H_1^0}$, ${\tilde H_2^0}$.  In
this basis, their mass matrix is given by
\begin{eqnarray} \label{eq:neumass}
   {\mathcal M}_{\tilde \chi^0_{1,2,3,4}} =
   \left( \matrix{
   {M_1} & 0 & -{g'v_1\over\sqrt{2}} & +{g'v_2\over\sqrt{2}} \cr
   0 & {M_2} & +{gv_1\over\sqrt{2}} & -{gv_2\over\sqrt{2}} \cr
   -{g'v_1\over\sqrt{2}} & +{gv_1\over\sqrt{2}} & \delta_{33} & -\mu
\cr
   +{g'v_2\over\sqrt{2}} & -{gv_2\over\sqrt{2}} & -\mu & \delta_{44}
\cr
   } \right)
\end{eqnarray}
where $g$ and $g'$ are the gauge coupling constants of SU(2) and U(1). 
$\delta_{33}$ and $\delta_{44}$ are the most important one-loop
corrections.  These can change the neutralino masses by a few GeV up
or down and are only important when there is a severe mass degeneracy
of the lightest neutralinos and/or charginos.  The expressions for
$\delta_{33}$ and $\delta_{44}$ used in \ds\ are taken from
\cite{NeuLoop1,NeuLoop2} (the tree-level values can optionally be
chosen).

The neutralino mass matrix, Eq.~(\ref{eq:neumass}), is
diagonalized analytically and evaluated numerically
to give the masses and compositions of four neutral Majorana states,
\begin{equation}
   \tilde{\chi}^0_i =
   N_{i1} \tilde{B} + N_{i2} \tilde{W}^3 +
   N_{i3} \tilde{H}^0_1 + N_{i4} \tilde{H}^0_2 ,
\end{equation}
the lightest of which, $\tilde\chi^0_1$ to be called $\chi$ for 
simplicity, is then the candidate for
the particle making up the dark matter in the Universe. The neutralinos are ordered in mass such that $m_{\chi_1^0} < m_{\chi_2^0} < m_{\chi_3^0} < m_{\chi_4^0}$ and the eigenvalues are real with a complex $N$. 

The charginos are linear combinations of the charged gauge bosons
${\tilde W^\pm}$ and of the charged Higgsinos ${\tilde H_1^-}$,
${\tilde H_2^+}$. Their mass terms are given by
\begin{equation}
   \left( \matrix{ {\tilde W^-} & {\tilde H_1^-} } \right)
   \> {\mathcal M}_{\tilde{\chi}^\pm} \>
   \left( \matrix{ {\tilde W^+} \cr {\tilde H_2^+} } \right)
   + \mbox{\rm h.c.}
\end{equation}
Their mass matrix,
\begin{eqnarray}
   {\mathcal M}_{\tilde{\chi}^\pm} =
   \left( \matrix{
   {M_2} & {gv_2} \cr
   {gv_1} & \mu
   } \right) ,
\end{eqnarray}
is diagonalized by the following linear combinations
\begin{eqnarray}
   \tilde{\chi}^-_i & = & U_{i1} \tilde{W}^- + U_{i2} \tilde{H}_1^- ,
\\
   \tilde{\chi}^+_i & = & V_{i1} \tilde{W}^+ + V_{i2} \tilde{H}_2^+ .
\end{eqnarray}
We choose ${\rm det}(U)=1$ and $U^* {\mathcal M}_{\tilde{\chi}^\pm}
V^\dagger = {\rm diag} ( m_{\tilde{\chi}^\pm_1},
m_{\tilde{\chi}^\pm_2} )$ with non-negative chargino masses $
m_{\tilde{\chi}^\pm_i} \ge 0$. 
We do not include any one-loop corrections to the chargino masses
since they are negligible compared to the corrections $\delta_{33}$
and $\delta_{44}$ introduced above for the neutralino masses
\cite{NeuLoop1}.

\subsection{Sfermion masses and mixings}
When discussing the squark mass matrix including mixing, it is
convenient to choose a basis where the squarks are rotated in the same
way as the corresponding quarks in the Standard Model.  We follow the
conventions of the Particle Data Group \cite{PDG} and put the mixing
in the left-handed $d$-quark fields, so that the definition of the
Cabibbo-Kobayashi-Maskawa matrix is $\mbox{\bf K}= \mbox{\bf V}_1
\mbox{\bf V}_2^\dagger$, where $\mbox{\bf V}_1$ ($\mbox{\bf V}_2$)
rotates the interaction left-handed $u$-quark ($d$-quark) fields to
mass eigenstates.  For sleptons we choose an analogous basis, but
since in \ds\ 4.1 neutrinos are assumed to be massless, no analog 
of the CKM matrix appears.

The general $6\times6$ $\tilde{u}$- and
$\tilde{d}$-squark mass matrices are
\begin{equation}
   {\mathcal M}_{\tilde u}^2 = \left( \matrix{
   \mbox{\bf M}_Q^2 + \mbox{\bf m}_u^\dagger \mbox{\bf m}_u +
       D_{LL}^{u} \mbox{\bf 1} &
    \mbox{\bf m}_u^\dagger
         ( {\bf A}_U^\dagger - \mu^* \cot\beta ) \cr
    ( {\bf A}_U - \mu \cot\beta ) \mbox{\bf m}_u &
   \mbox{\bf M}_U^2 + \mbox{\bf m}_u \mbox{\bf m}_u^\dagger +
       D_{RR}^{u} \mbox{\bf 1} \cr
   } \right),
   \label{mutilde}
\end{equation}
\begin{equation}
   {\mathcal M}_{\tilde d}^2=\left( {\matrix{
   {\mbox{\bf K}^\dagger \mbox{\bf M}_Q^2 \mbox{\bf K}+
   \mbox{\bf m}_d\mbox{\bf m}_d^\dagger+D_{LL}^{d}\mbox{\bf 1}}&
   {\mbox{\bf m}_d^\dagger ( {\bf A}_D^\dagger-\mu^*\tan\beta )}\cr
   {( {\bf A}_D-\mu\tan\beta ) \mbox{\bf m}_d}&
   {\mbox{\bf M}_D^2+\mbox{\bf m}_d^\dagger\mbox{\bf m}_d+
       D_{RR}^{d}\mbox{\bf 1}}\cr
   }} \right),
   \label{mdtilde}
\end{equation}
The general sneutrino and charged slepton mass matrices are (for massless neutrinos)
\begin{equation}
   {\mathcal M}^2_{\tilde\nu} = \mbox{\bf M}_L^2 + D^\nu_{LL} \mbox{\bf 1}
\end{equation}
\begin{equation}
   {\mathcal M}^2_{\tilde e} =\left( {\matrix{
   {\mbox{\bf M}_L^2+\mbox{\bf m}_e\mbox{\bf m}_e^\dagger+
        D_{LL}^{e}\mbox{\bf 1}}&
   {\mbox{\bf m}_e^\dagger ( {\bf A}_E^\dagger-\mu^*\tan\beta )}\cr
   {( {\bf A}_E-\mu\tan\beta ) \mbox{\bf m}_e}&
   {\mbox{\bf M}_E^2+\mbox{\bf m}_e^\dagger\mbox{\bf m}_e+
        D_{RR}^{e}\mbox{\bf 1}}\cr
   }} \right).
   \label{metilde}
\end{equation}
Here
\begin{equation}
   D^f_{LL}=m_Z^2\cos 2\beta(T_{3f}-e_f\sin^2\theta_W),
\end{equation}
\begin{equation}
   D^f_{RR}=m_Z^2\cos(2\beta) e_f\sin^2\theta_W
\end{equation}
where $T_{3f}$ is the third component of the weak isospin and $e_{f}$
is the charge in units of the absolute value of the electron charge, $e$.
In the chosen basis, we have $\mbox{\bf m}_u$ = $\mbox{\rm diag}
\left( m_{\rm
u}, m_{\rm c}, m_{\rm t} \right)$, $\mbox{\bf m}_d $ = $\mbox{\rm
diag}
\left(m_{\rm d}, m_{\rm s}, m_{\rm b} \right)$ and $\mbox{\bf m}_e $
= $
\mbox{\rm diag} (m_e, m_\mu, m_\tau )$.

The slepton and squark mass eigenstates $\tilde{f}_k$ ($\tilde{\nu}_k$
with $k=1,2,3$ and $\tilde{e}_k$, $\tilde{u}_k$ and $\tilde{d}_k$ with
$k=1,\dots,6$) diagonalize the previous mass matrices and are related
to the current sfermion eigenstates $\tilde{f}_{La}$ and
$\tilde{f}_{Ra}$ ($a=1,2,3$) via
\begin{eqnarray}
   \tilde{f}_{La} & = & \sum_{k=1}^6 \tilde{f}_k {\bf
\Gamma}_{FL}^{*ka} , \\
   \tilde{f}_{Ra} & = & \sum_{k=1}^6 \tilde{f}_k {\bf
\Gamma}_{FR}^{*ka} .
\end{eqnarray}
The squark and charged slepton mixing matrices ${\bf \Gamma}_{UL}$, 
${\bf \Gamma}_{UR}$,
${\bf \Gamma}_{DL}$, ${\bf \Gamma}_{DR}$, ${\bf \Gamma}_{EL}$,
 and ${\bf \Gamma}_{ER}$ have dimension
$6\times 3$, while the sneutrino mixing matrix ${\bf \Gamma}_{\nu L}$
has dimension $3\times3$.

The current version of \ds\ allows only for diagonal matrices ${\bf
A}_U$, ${\bf A}_D$, ${\bf A}_E$, ${\bf M}_Q$, ${\bf M}_U$, ${\bf
M}_D$, ${\bf M}_E$, and ${\bf M}_L$.  This ansatz, while not being the
most general, implies the absence of tree-level flavor changing
neutral currents in all sectors of the model.  It also allows the squark
mass matrices to be diagonalized analytically.  For example, for the
top squark one has, in terms of the top squark mixing angle
$\theta_{\tilde{t}}$,
\begin{equation}
   \Gamma_{UL}^{\tilde{t}_1\tilde{t}} =
   \Gamma_{UR}^{\tilde{t}_2\tilde{t}} = \cos \theta_{\tilde{t}} ,
   \qquad
   \Gamma_{UL}^{\tilde{t}_2\tilde{t}} =
   - \Gamma_{UR}^{\tilde{t}_1\tilde{t}} = \sin \theta_{\tilde{t}} .
\end{equation}
The sfermion masses are obtained with the diagonalization just described.

To facilitate comparisons with the results of other authors, \ds\ allows for special
values of the sfermion masses to be set in the program.  A common value can be
assigned to all squark masses, and an independent common value to all slepton
masses. Alternatively, to enforce the sfermions to be heavier than the lightest
neutralino (which we want to be the LSP), the squarks and sleptons masses can be
set equal to the maximum between the neutralino mass and a specified value.  In
the special cases just described, no mixing is assumed between sfermions. It
must be noted that the special choices described in this paragraph are
mathematically inconsistent within the MSSM, but are often made for the sake of
simplicity.

\subsection{Higgs sector and interface to FeynHiggs}

The Higgs masses receive radiative corrections and \ds\ includes several 
options for calculating these.  The default in \ds\ is to use \code{FeynHiggsFast} \cite{Heinemeyer:2000nz} for the Higgs mass calculations. When higher accuracy is needed, it is possible to instead use the full \code{FeynHiggs} \cite{feynhiggs1,feynhiggs2} package.

\subsection{Interface to the mSUGRA codes ISASUGRA and SUSPECT}

Given the modular structure of \ds, the user may also run the package using as
input for the MSSM definition the output, still at the low energy scale, from an external
package. An example of usage under such mode is given in case of the minimal 
supergravity (mSUGRA) model: in the release, we provide an interface to the
output of the \code{ISASUGRA} code, as included in ISAJET \cite{isajet} for the 
current \code{ISASUGRA 7.69} version, as well as an an interface to 
\code{SUSPECT} \cite{suspect} which can be used as an alternative.
The interfaces to \ds\ are at the level of the full spectrum of masses and mixings, 
including, for consistency, those for the Higgs sector. Of importance for relic density calculations near a Higgs boson resonance is the possibility of
including supersymmetric corrections to the bottom, top, and tau Yukawa couplings as supplied by \code{ISASUGRA}.

\section{Experimental constraints}

Accelerator bounds can be checked by a call to a subroutine. By modifying an
option, the user can impose bounds as of different moments in time. The default
option in version \dsver\ adopts the 2002 limits by the 
Particle Data Group
\cite{PDG02} modified as described below. The user is also free to use his or her 
own routine to check for experimental bounds, in which case there is
only need to provide an interface to \ds.

For the theoretical prediction of the rare decay $b \rightarrow s \gamma$ we have implemented the complete next-to-leading order (NLO) Standard Model calculation and the dominant NLO supersymmetric corrections. For the NLO QCD calculation of the Standard Model prediction we have used the expressions of reference \cite{bsgsm} into which we have inserted the updated so-called ``Magic numbers'' of  \cite{bsgmagic}. Our implementation of the Standard Model calculation gives a branching ratio $\mathrm{BR}[B\rightarrow X_s\,\gamma] =3.72\times10^{-4} $ for a photon energy greater than  $m_b/20$. This result agrees to within 1\% with the result stated in \cite{bsgmagic}, but is around 10\% larger than the result of previous analyses. The latter is due to the fact that in the reference \cite{bsgsm} they replaced $m_c^{pole}/m_b^{pole}$ by $m_c^{\overline{MS}}(\mu)/m_b^{pole}$ (with $\mu\in [m_c,m_b]$) in the matrix element \mbox{$\langle X_s\gamma\mid(\bar{s}c)_{V-A}(\bar{c}b)_{V-A}\mid{b}\rangle$}. 

The supersymmetric correction to $b \rightarrow s \gamma$ has been divided into a contribution from a two Higgs doublet model and a contribution from supersymmetric particles. The expressions for the NLO contributions in the two Higgs doublet model has been taken from \cite{bsgh2}. The corrections due to supersymmetric particles are calculated under the assumption of minimal flavour violation. The dominant LO contributions which are valid even in the large $\tan\beta$ regime was taken from ref.~\cite{bsgtan}, and we also followed their guideline on how the NLO QCD expressions of \cite{bsgsusy} should be expanded to the large $\tan\beta$ regime.

For the current experimental bound on $b \rightarrow s \gamma$ we take the value stated by the Particle Data Group 2002 \cite{PDG02}. This is an average between the CLEO and the Belle measurements and amounts to $\mathrm{BR}[B\rightarrow X_s\,\gamma] =(3.3\pm0.4)\times10^{-4}$. To this we add a theoretical uncertainty which we set to $\pm0.5\times10^{-4}$. The final constraint on the branching ratio then becomes $2.0\times10^{-4}\leq\mathrm{BR}[B\rightarrow X_s\,\gamma] \leq4.6\times10^{-4}$.

Recently, much interest has been given to the possible contribution
of supersymmetry to $(g-2)_\mu$. Although the discrepancy with the 
Standard Model result is now below 3$\sigma$, we include for
convenience
 a calculation of the anomalous moment of the muon $(g-2)_\mu$ in \ds.


\section{Calculation of relic density}
\label{RelDens}

The WMAP microwave background experiment \cite{wmap1}, 
combined with
other sets of data, gives a quite precise determination of the cold dark matter density
$\Omega_{CDM} h^2 = 0.113 \pm 0.009$. We would like \ds\ to 
compute the relic density of neutralinos to at least the same precision. 

We use in \ds\ the full cross section and
solve the Boltzmann equation numerically with the method given
in~\cite{GondoloGelmini,coann}. In this way we automatically take
care of thresholds and resonances.

When any other supersymmetric particles are close in mass to the
lightest neutralino they will also be present at the time when the
neutralino freezes out in the early Universe.  When this happens so-called 
coannihilations can take place between all these supersymmetric
particles present at freeze-out.  
Coannihilations were first pointed out by Binetruy, Girardi, and Salati \cite{binetruy} in a non-supersymmetric model with several Higgs bosons. Griest and
Seckel \cite{griestseckel} investigated them in the MSSM for the 
 case where squarks are of about the same mass as the
lightest neutralino.  Later, coannihilations between the lightest
neutralino and the lightest chargino were investigated in
\cite{MizutaYamaguchi, DreesNojiri, NeuLoop1}. Several authors have also included coannihilations with sfermions\cite{Ellisstau1,Ellisstau2,Ellisstop,Ellislast,Ellislargetg,
BDD,GLP,micromega,bbb,roszkowski,GLP2,BKKZ,BKKG,bbb7.64,ADS,coann2}. In DarkSUSY we have implemented all
coannihilations between the neutralinos, charginos and sfermions as calculated in \cite{coann2}.

Compared to other recent calculations, we believe ours is the most precise calculation available at present. The standard lore so far
has been to calculate the thermal average of the
annihilation cross section by
expanding to first power in temperature over mass and implementing
an approximate solution to the evolution equation which estimates the 
freeze out temperature without fully solving the equation
(see, e.g., Kolb and Turner~\cite{KT}). Sometimes this is refined by
including resonances and threshold corrections~\cite{griestseckel}.
Among recent studies, this approach is taken in e.g.~Refs.~\cite{BDD,GLP}.
Other refinements include, e.g., solving the density evolution equation
numerically but still using an approximation to thermal effects in the
cross section~\cite{Ellisstau1,Ellisstau2,Ellisstop,Ellislast,Ellislargetg},
or calculating the thermal average precisely but using an
approximate solution to the density 
equation~\cite{micromega,bbb,roszkowski}. At the same time, only in a 
few of the quoted papers the full set of initial states has been included.
As already mentioned, the present calculation includes all initial states,
performs an accurate thermal average and gives a very precise solution to 
the evolution equation. Though the inclusion of initial state sfermions in 
the \ds\ package is a new feature introduced in the present work, other 
groups~\cite{GLP2,BKKZ,BKKG} have earlier introduced some sfermion coannihilations 
in an interface with the old \ds\ version.

To gain calculational speed we only include the particles with masses below
$f_{co}m_{\chi}$. The mass fraction
parameter $f_{co}$ is by default set to 2.1 or 1.4 depending on how
the relic density routines are called (very high precision or fast
calculation), but can be set to any value by the user.

\subsection{The Boltzmann equation}

We will here outline the procedure developed in \cite{coann} which is
used in \ds. For more details, see \cite{coann}.

Consider annihilation of $N$ supersymmetric particles with masses
$m_{i}$ and internal degrees of freedom $g_{i}$.  Order them such that
$m_{1} \leq m_{2} \leq \cdots \leq m_{N-1} \leq m_{N}$. For the
lightest neutralino, the notation $m_1$ and $m_\chi$ will be used
interchangeably.

Since we assume that $R$-parity holds, all supersymmetric particles will
eventually decay to the LSP and we thus only have to consider the
total number density of supersymmetric particles $ n= \sum_{i=1}^N n_{i}$.
Furthermore, the scattering rate of supersymmetric particles off 
particles in the
thermal background is much faster than their annihilation rate,
because the scattering cross sections $\sigma'_{Xij}$ are of the same
order of magnitude as the annihilation cross sections $\sigma_{ij}$
but the background particle density $n_X$ is much larger than each of
the supersymmetric particle densities $n_i$ when the former are
relativistic and the latter are non-relativistic, and so suppressed by
a Boltzmann factor \cite{thermal-freeze-out}.
Hence, the $\chi_i$ distributions remain in
kinetic thermal equilibrium during their freeze-out. Combining these effects,
we arrive at the following Boltzmann equation for the summed number
density of supersymmetric particles
\begin{equation} \label{eq:Boltzmann2}
   \frac{dn}{dt} =
   -3Hn - \langle \sigma_{\rm{eff}} v \rangle
   \left( n^2 - n_{\rm{eq}}^2 \right)
\end{equation}
where
\begin{equation} \label{eq:sigmaveffdef}
   \langle \sigma_{\rm{eff}} v \rangle = \sum_{ij} \langle
   \sigma_{ij}v_{ij} \rangle \frac{n_{i}^{\rm{eq}}}{n^{\rm{eq}}}
   \frac{n_{j}^{\rm{eq}}}{n^{\rm{eq}}}.
\end{equation}
with
\begin{equation}
   v_{ij} = \frac{\sqrt{(p_{i} \cdot p_{j})^2-m_{i}^2 m_{j}^2}}{E_{i} E_{j}}.
\end{equation}

\subsection{Thermal averaging}
\label{sec:thermav}

Using the Maxwell-Boltzmann approximation for the velocity
distributions one can derive the following expression for the
thermally averaged annihilation cross section \cite{coann}
\begin{equation} \label{eq:sigmavefffin2}
   \langle \sigma_{\rm{eff}}v \rangle = \frac{\int_0^\infty
   dp_{\rm{eff}} p_{\rm{eff}}^2 W_{\rm{eff}} K_1 \left(
   \frac{\sqrt{s}}{T} \right) } { m_1^4 T \left[ \sum_i \frac{g_i}{g_1}
   \frac{m_i^2}{m_1^2} K_2 \left(\frac{m_i}{T}\right) \right]^2}\,.
\end{equation} where $K_1$ ($K_2$) is the modified Bessel function of the second kind of order 1 (2), $T$ is the temperature, $s$ is one of the Mandelstam variables and
\begin{eqnarray} \label{eq:weff}
   W_{\rm{eff}} & = & \sum_{ij}\frac{p_{ij}}{p_{\rm{eff}}}
   \frac{g_ig_j}{g_1^2} W_{ij} \nonumber \\
   & = &
   \sum_{ij} \sqrt{\frac{[s-(m_{i}-m_{j})^2][s-(m_{i}+m_{j})^2]}
   {s(s-4m_1^2)}} \frac{g_ig_j}{g_1^2} W_{ij}.
\end{eqnarray}
In this equation, we have defined the momentum
\begin{equation}
    p_{ij} =
   \frac{\left[s-(m_i+m_j)^2\right]^{1/2}
   \left[s-(m_i-m_j)^2\right]^{1/2}}{2\sqrt{s}},
\end{equation}
the invariant annihilation rate
\begin{equation} \label{eq:Wijcross}
   W_{ij} = 4 p_{ij} \sqrt{s} \sigma_{ij} = 4 \sigma_{ij} \sqrt{(p_i
\cdot p_j)^2 - m_i^2 m_j^2} = 4 E_{i} E_{j} \sigma_{ij} v_{ij}
\end{equation}
and the effective momentum
\begin{equation} \label{eq:peff}
    p_{\rm{eff}} = p_{11} = \frac{1}{2}
   \sqrt{s-4m_{1}^2}.
\end{equation}
Since $W_{ij}(s) = 0 $ for $s \le (m_i+m_j)^2$, the radicand in
Eq.~(\ref{eq:weff}) is never negative.
For a two-body final state, $W_{ij}$ is given by
\begin{equation} \label{eq:Wij2body}
   W^{\rm{2-body}}_{ij} = \frac{|\bf k|}{16\pi^2 g_i g_j S_f \sqrt{s}}
   \sum_{\rm{internal~d.o.f.}} \int \left| {\cal M} \right|^2
   d\Omega ,
\end{equation}
where $\bf k$ is the final center-of-mass momentum, $S_f$ is a symmetry
factor equal to 2 for identical final particles, and the integration
is over all possible outgoing directions of one of the final particles.
As usual, an average over initial internal degrees of freedom is
performed.

\subsection{Annihilation cross sections}
\label{sec:AnnChannels}

In \ds,  all two-body final state cross sections at tree
level are computed for all coannihilations between neutralinos, charginos and sfermions.
A complete list is given in table~\ref{tab:coanns} in Appendix \ref{sec:processes}.

The calculation of the relic density is, due to its complexity, the
most time-consuming task of \ds.  For the neutralino-neutralino,
chargino-neutralino and chargino-chargino initial states,
to achieve efficient numerical
computation, contributing diagrams have been classified according to
their topology ($s$-, $t$- or $u$-channel) and to the spin of the
particles involved.  The helicity amplitudes for each type of diagram
have been computed analytically with \code{Reduce}~\cite{reduce} using
general expressions for the vertex couplings.  The \code{Reduce} output
has been automatically converted to \code{FORTRAN} subroutines called
by \ds.

The strength of the helicity amplitude method is that the analytical
calculation of a given diagram only has to be performed once and the
summing of the contributing diagrams for each given set of initial and
final states can be done numerically afterwards.

The Feynman diagrams are summed numerically for each
annihilation channel $ij\to kl$. We then sum the squares of the
helicity amplitudes and sum the contributions of all
annihilation channels. Explicitly, \ds\ computes
\begin{equation} \label{eq:helsum}
   {d W_{\rm eff} \over d \cos\theta } =
\sum_{ijkl}
{p_{ij} p_{kl} \over 32 \pi S_{kl} \sqrt{s} }
\sum_{\rm helicities}
    \left| \sum_{\rm diagrams}  {\cal M}(ij \to kl) \right|^2
\end{equation}
where $\theta$ is the angle between particles $k$ and $i$.  Finally,
a numerical integration over $\cos\theta$ is performed
by means of an adaptive method \cite{dqagse}. 

In rare cases, there can be resonances in the $t$- or $u$-channels.
This can happen when the masses of the initial state particles lie between 
the masses of the final
state particles. At certain values of $\cos\theta$, the momentum transfer is
time-like and matches the mass of the exchanged particle.  We regulate
the divergence by assigning a small width of 5 GeV to the neutralinos
and charginos and 1 GeV to the sfermions.  The results are not
sensitive to the exact choice of this width.

For the coannihilation diagrams with sfermions, the calculations are
done with \code{Form} \cite{form} and automatically converted into {\tt
FORTRAN} subroutines.

In the relic density routines, the calculation of the effective
invariant rate $W_{\rm eff}$ is the most time-consuming part.
Fortunately, thanks to the remarkable feature of
Eq.~(\ref{eq:sigmavefffin2}), $W_{\rm eff}(p_{\rm eff})$ does not
depend on the temperature $T$, and it can be tabulated once for each
model.

To perform the thermal average in Eq.~(\ref{eq:sigmavefffin2}), we
integrate over $p_{\rm eff}$ by means of adaptive gaussian
integration, using a spline routine to interpolate in the $(p_{\rm
eff},W_{\rm eff})$ table. To avoid numerical problems in the
integration routine or in the spline routine, we split the integration
interval at each sharp threshold. We also explicitly check for each
MSSM model that the spline routine behaves well close to thresholds and
resonances.

We then integrate the density evolution equation, Eq.~(\ref{eq:Boltzmann2}). For numerical reasons, we do not integrate the equation directly, but instead rephrase it as an evolution equation for the abundance, $Y=n/s$ (with $s$ being the entropy density) and use $x=m_\chi/T$ as our integration variable instead of time (see \cite{coann} for details). The numerical integration 
is subtle since the equation is ``stiff.'' 
For this purpose, we developed an implicit trapezoidal method with adaptive stepsize.
In short, if the equation for $Y$ is written as $dY/dx = \lambda (Y^2-Y_e^2)$, with $Y_e$ being the equilibrium value at temperature $T$, the numerical integration is based on the recurrence
\begin{equation} 
Y_{i+1} = \frac{ C_i }{ 1 + \sqrt{1+h \lambda_{i+1} C_i} }
\end{equation}
where $h$ is the stepsize and $ C_i = 2 Y_i + h [ \lambda_{i+1} Y_{e,i+1}^2 + \lambda_i (Y_{e,i}^2 - Y_i^2) ] $. The stepsize is adapted (reduced) if $| (Y_{i+1}-y_{i+1})/Y_{i+1}|$ exceeds a given tolerance. Here 
$ y_{i+1} = (c_i/2) ( 1 + \sqrt{1+h \lambda_{i+1} c_i} )^{-1} $
with $ c_i = 4 ( Y_i +  h \lambda_i Y_{e,i}^2 ) $. 

There are some loop-induced final states, such as two gluons, two
photons or a $Z^0$ boson and a photon which could in principle give a
non-negligible contribution to the annihilation rate and lower the
relic abundance somewhat.  The cross sections for these
processes are very complicated already in the limit of zero velocity,
and we therefore assume that the invariant rate $W$ for these one-loop
processes are constant and equal to their zero-momentum expressions.
These processes can be excluded from the calculation
by setting appropriate parameters in the code.

The relic density routines can be called in a precise way where all
integrations are performed to a precision better than 1\% and coannihilations
are included up to a mass difference of $f_{co}=2.1$. It can also be
called in a faster way, where the precision of the integrations are of
the order of 1\% and coannihilations are included up to $f_{co}=1.4$.
Usually, the difference between the precise and fast method is
completely negligible, but in rare cases it can be of a few \%. The fast
option is considerably faster and should be sufficient for most
purposes. For advanced users, it is also possible to manually decide exactly which coannihilation processes to include.

For users with less demand of calculational precision, we also provide
in \ds\ the option of a ``quick-and-dirty'' method of computing the
relic density, essentially according to the textbook treatment
in \cite{KT}. It should be realised, however, that this
may sometimes give a computed relic density which is wrong
by orders of magnitude.

\subsection{A note about degrees of freedom and the annihilation rate}\label{subs:note}

We end this section with a comment on the internal degrees of
freedom $g_i$.  A neutralino is a Majorana fermion (it is its
own antiparticle) and has two
internal degrees of freedom, $g_{\chi_{i}^0}=2$. A chargino can be
treated either as two separate species $\chi_{i}^+$ and
$\chi_{i}^-$, each with internal degrees of freedom
$g_{\chi^+}=g_{\chi^-}=2$, or, more simply, as a single species
$\chi_{i}^\pm$ with $g_{\chi_{i}^\pm}=4$ internal degrees of
freedom. We choose the latter convention in \ds\ and use the analogous conventions for sfermions (see \cite{coann2} for details).

The counting of states in annihilation processes for Majorana
fermions is non-trivial,
and has led to a factor-of-two ambiguity in the literature 
which also propagated into earlier versions of \ds. The current
version of \ds\ contains the correct normalization
of annihilation rates, namely $\sigma v$ in the Boltzmann equation and $\sigma v/2$ for annihilation in the halo. 
The clearest way to see the origin of the factor of 
$1/2$ is probably to go back to the Boltzmann equation, \cite{BEU}.
In essence, one can view $\sigma v$ as the thermal average (averaged over 
momentum and angles) of the cross section times velocity in the zero momentum 
limit; in this average one integrates over all possible angles.
For identical particles in the initial state, one includes each possible 
initial state twice, therefore one needs to compensate by dividing
by a factor of 2; 
the prefactor in the zero-momentum limit
becomes then $\sigma v/2$. In the Boltzmann equation describing the
time evolution of the neutralino number density the $1/2$ does not appear
as it is compensated by the factor of 2 one has to include because 2
neutralinos are depleted per annihilation, but we need to include the factor of $1/2$ explicitly in other cases where we need the annihilation rate (like for annihilation in the halo).
Another way to view this is to think of $\sigma$ as to the annihilation
cross section for a given pair of particles. Let the number of neutralinos in a given
volume be $N$; the annihilation rate would be given by $\sigma v$ times the 
number of pairs, which is $N(N-1)/2$. In the continuum limit this reduces to 
$\sigma v n^2 /2$.

\section{Halo models}
\label{sec:halo}

The modelling of the distribution of dark matter particles in the Milky
Way dark halo plays a major role in estimates of dark matter detection 
rates. On this issue, however, there is no well-established framework we 
can refer to. Available dynamical measurements, such as, e.g., the mapping 
of the rotation curve, the local field of acceleration of stars in the 
direction perpendicular to the disk, or the motion of the satellites in 
the outskirts of the Galaxy, provide some constraints on the dark matter 
density profile, but lack the precision needed to derive a refined model. 
The guideline for the future will be N-body simulations of hierarchical 
clustering in cold dark matter cosmologies, which are starting to resolve 
the inner structure of individual galactic halos. At present, however, the 
translation of such results into the detailed model we need for the Milky 
Way still relies on large extrapolations, and, to some extent, faces the 
problem of possible discrepancies between some of its properties and 
observations in real galaxies (for a recent review see, 
e.g.,~\cite{primackrev}).

In light of these large uncertainties, the definition of the model for the 
dark matter halo in \ds\ has been kept in a very general format. Two sets 
of properties need to be implemented:

\begin{itemize}
     \item[a)] Local properties: To derive counting rates in direct 
   detection and $\chi$-induced neutrino fluxes from the center of the Sun 
   and the Earth, the user has to specify the halo mass density $\rho_{0}$ 
   at our galactocentric distance $R_0$ and the relative particle velocity 
   distribution $f(u)$, where $u$ is the modulus of the velocity $\vec{u}$ 
   in the rest frame of the Sun (or Earth, respectively) and an average of incident 
   angles has been assumed. Options to set $f(u)$ include the possibility 
   to implement the expression valid for a (truncated) isothermal sphere, 
   or the numerical interpolation of a table of pairs $(f_i, u_i)$ provided 
   by the user. We have also implemented routines which compute the function
   $f(u)$ in the Earth or the Sun rest frame for a given isotropic velocity 
   distribution function in the Galactic reference frame as needed for 
   direct detection and for neutralino capture rates by the Sun. Finally
   a numerical table for $f(u)$ in the Earth frame and including the 
   modelling of the diffusion process of neutralinos through the solar 
   system~\cite{le} is available to compute capture rates by the Earth. 

     \item[b)] Global properties: To compute indirect signals from pair 
   annihilations in the halo, the full dark matter mass density profile is 
   needed. Charged cosmic ray fluxes are computed in two dimensional models 
   and a generic axisymmetric profile $\rho(R,z)$ can be implemented by the 
   user. Line of sight integrals with angular averaging over acceptance 
   angles, needed to compute gamma-ray fluxes, are given for spherically 
   symmetric profiles $\rho(r)$. The options to specify $\rho$ include 
   several analytic forms, with most classes of profiles proposed in recent 
   years~\cite{navarro,moore,n03,burkert}, as well as the numerical 
   interpolation of a table of pairs 
   $(\rho_i,r_i)$ provided by the user. It is also possible to switch on an 
   option to compute the signals for annihilations taking place in a 
   population of small, unresolved clumps. In this case the user should 
   specify the probability distribution for the clumps and the average 
   enhancement of the source per unit volume compared to the smooth halo 
   scenario.
\end{itemize}

An option to rescale $\rho_0$ and $\rho(r)$ by the quantity 
$\Omega h^2 / \Omega_{\rm min} h^2$, where $\Omega h^2$ is the neutralino 
relic abundance and $\Omega_{\rm min} h^2$ a minimum reference value for 
neutralinos to provide most of the dark matter in our Galaxy, is available for the  
case where subdominant dark matter candidates are considered. The effect can 
be introduced a posteriori as well for all detection techniques except for 
neutrino fluxes, where the response to rescaling is non-linear. 

A separate package interfaced to \ds\ with self consistent pairs 
$(\rho(r), f(u))$ in $\Lambda$CDM inspired models which fit available 
dynamical constraints will be available shortly from one of the 
authors~\cite{ulliocdm}.


\section{Detection rates}
For the detection rates of neutralino dark matter we have used the
rates as calculated in Refs.\
\cite{BEU,lpj,bg,beg,pbar,baltz,eg,joakimthesis,dkpop}, with some
improvements which we report in this Section where the formulas used
in \ds\ are presented.

\subsection{Direct detection}\label{subs:direct}

The current version of \ds\ provides the neutralino-proton and 
neutralino-neutron scattering cross sections (spin-independent and spin-dependent), as well as cross sections and form factors for elastic scattering of neutralinos off nuclei.

The rate for direct detection of galactic neutralinos can be written as
\begin{equation}
\frac{dR}{dE} = \sum_i c_i \frac{ \rho_\chi \sigma_{\chi i} |F_i(q_i)|^2}{2 m_\chi \mu^2_{i\chi} } \int_{v>\sqrt{M_iE/2\mu^2_{\chi i}}} \frac{f({\bf v},t)}{v} d^3 v . 
\end{equation}
The sum is over the nuclear species in the detector, $c_i$ being the detector mass fraction in nuclear species $i$. $M_i$ is the nuclear mass, and $\mu_{\chi i} = m_\chi M_i /(m_\chi + M_i)$ is the reduced neutralino--nucleus mass. Moreover, $\rho_\chi$ is the local neutralino density, ${\bf v}$ the neutralino velocity relative to the detector, $v=|{\bf v}|$, and $f({\bf v},t) d^3v$ the neutralino velocity distribution. Finally,  $\sigma_{\chi i}$ 
is the total scattering cross section of a WIMP off a fictitious point-like nucleus, $|F_i(q_i)|^2$ is a nuclear form factor that depends on the momentum transfer $q_i=\sqrt{2M_iE}$ and is normalized to $F_i(0)=1$. The integration is over all neutralino speeds that can impart a recoil energy $E$ to the nucleus. 

The cross section $\sigma_{\chi i}$ scales differently for spin-dependent and spin-independent interactions. 
For spin-independent interactions with a nucleus with $Z$ protons and $A-Z$ neutrons, one has
\begin{equation}
\label{eq:spinindependent}
  \sigma_{\chi i}^{\rm SI} = \frac{\mu^2_{\chi i}}{\pi} \big| Z G^p_s + (A-Z) G^n_s \big|^2 ,
\end{equation}
where $G^p_s$ and $G^n_s$ are the scalar four-fermion couplings of the WIMP
with point-like protons and neutrons, respectively (see, e.g., \cite{lesarcs}, and below). As default, the spin-independent form factor in \ds\ is taken to be of the Helm form
\begin{equation}
|F^{\rm SI}(q)|^2 = \left( \frac{ 3 j_1(q R_1)}{q R_1} \right)^2 e^{-q^2 s^2},
\end{equation}
with $j_1$ a spherical Bessel function of first kind,
$R_1 = \sqrt{ R^2-5s^2}$, $R = [ 0.9 (M/{\rm GeV})^{1/3} +0.3 ] {\rm ~fm}$ and $s=1$ fm. An exponential form factor is also available as an option.

For spin-dependent interactions, one has instead, at zero momentum transfer,
\begin{equation}
  \sigma_{\chi i}^{\rm SD} = \frac{4\mu^2_{\chi i}}{\pi} \frac{J+1}{J} \big| \langle S_p \rangle G^p_a + \langle S_n \rangle G^n_a \big|^2 ,
  \end{equation}
  where $J$ is the  nuclear spin, $\langle S_p \rangle$ and $ \langle S_n \rangle$ are the expectation values of the spin of the protons and neutrons in the nucleus, respectively, and $G^p_a$ and $G^n_a$ are the axial four-fermion couplings of the WIMP
with point-like protons and neutrons \cite{lesarcs,tovey}. At finite momentum transfer, the spin-dependent cross section times the form factor reads
\begin{equation}
\sigma_{\chi i}^{SD} | F_i^{\rm SD}(q)|^2 = \frac{ 4 \mu^2_{\chi i}}{2J+1} \left[ (G^p_a)^2 S_{pp}(q) + (G^n_a)^2 S_{nn}(q) + G^p_a G^n_a S_{pn}(q) \right],
\end{equation}
where
$S_{pp}(q) = S_{00}(q)+S_{11}(q)+S_{01}(q)$,
$S_{nn}(q) = S_{00}(q)+S_{11}(q)-S_{01}(q)$,
and $S_{pn}(q) = 2[S_{00}(q)-S_{11}(q)]$.
The spin structure functions $S_{00}(q)$, $S_{11}(q)$, and $S_{01}(q)$ are defined in \cite{EngelVogel} and are given in the literature.

For protons and neutrons, the previous expressions reduce to
\begin{equation}
\sigma_{\chi p}^{\rm SI} = \frac{\mu_{\chi p}^2}{\pi} | G_s^p |^2, \quad 
\sigma_{\chi n}^{\rm SI} = \frac{\mu_{\chi n}^2}{\pi} | G_s^n |^2, \quad 
\sigma_{\chi p}^{\rm SD} = \frac{3\mu_{\chi p}^2}{\pi} | G_a^p |^2, \quad 
\sigma_{\chi n}^{\rm SD} = \frac{3\mu_{\chi n}^2}{\pi} | G_a^n |^2.
\end{equation}

For the neutralino, the scalar and axial four-fermion couplings with the proton and neutron arise from squark, Higgs and Z boson exchange. In \ds, the default expressions for a nucleon $N=n,p$ are
\begin{eqnarray}
G_s^N & = & \sum_{q=u,d,s,c,b,t} \langle N | \overline{q} q | N \rangle \left( \frac{1}{2} \sum_{k=1}^{6} \frac{ g_{L\tilde{q}_k \chi q} g_{R \tilde{q}_k \chi q} } { m_{\tilde{q}_k}^2 } - \sum_{h=H_1,H_2} \frac{ g_{h\chi\chi} g_{hqq}} {m_h^2 } \right) , \\
G_a^N & = & \sum_{q=u,d,s} (\Delta q)_N \left( \frac{g_{Z\chi\chi} g_{Zqq}}{m_Z^2} + \frac{1}{8} \sum_{k=1}^{6} \frac{ g_{L\tilde{q}_k \chi q}^2 + g_{R\tilde{q}_k \chi q}^2} { m_{\tilde{q}_k}^2 } \right),
\end{eqnarray}
where $g_{abc}$ are the coupling constants in the vertex involving particles $abc$ (see \cite{bg} and \cite{joakimthesis} for explicit expressions). The more complicated expressions of Drees and Nojiri \cite{dn} are also available as an option.

Default values of the parameters in \ds\ are \cite{Gasser,SMC} (with $\langle N | \overline{q} q | N \rangle = f^N_{Tq} m_N/m_q$) 
\begin{equation}
f^p_{Tu} = 0.023, \quad f^p_{Td} = 0.034,  \quad  f^p_{Ts} = 0.14,
 \quad f^p_{Tc} = f^p_{Tb}= f^p_{Tt} = 0.0595, 
 \end{equation}
 \begin{equation}
 f^n_{Tu} = 0.019, \quad f^n_{Td} = 0.041,  \quad f^n_{Ts} = 0.14, \quad
f^n_{Tc} = f^n_{Tb}= f^n_{Tt} = 0.0592.
\end{equation}
\begin{equation}
(\Delta u)_p = (\Delta d)_n = 0.77, \quad 
(\Delta d)_p = (\Delta u)_n = -0.40, \quad 
(\Delta s)_p = (\Delta s)_n = -0.12.
\end{equation}
Other sets of values for these parameters are available. These values can also be overridden by the user.

\subsection{Indirect detection}
There are several ways of detecting dark matter particles indirectly.
Pair annihilations of dark matter particles in the Galactic halo may give
an exotic component in positron, antiproton or antideuteron cosmic-rays
and gamma-rays. There may also be  annihilation
in astrophysical environments where the density of neutralinos 
may be enhanced,
such as annihilation in the center of the Earth or Sun (detected 
in neutrino telescopes through a high-energy neutrino flux) or near the
central black hole of the Galaxy.
\subsubsection{Monte Carlo simulations}
\label{sec:mcsim}

In several of the indirect detection processes below we need to
evaluate the yield of different particles per neutralino annihilation.
The hadronization and/or decay of the annihilation products are
simulated with \code{Pythia} \cite{pythia} 6.154
in essentially the same way (with a few differences) for all
these processes and we here describe how the simulations are done.
We can divide the simulations into two groups: a) annihilation in the
Earth and the Sun and b) annihilation in the halo.
In both cases the simulations are done for a set of 18 neutralino
masses, $m_{\chi}$ = 10, 25, 50, 80.3, 91.2, 100, 150, 176, 200, 250,
350, 500, 750, 1000, 1500, 2000, 3000 and 5000 GeV\@. We tabulate the
yields and then interpolate these tables in \ds.

These two groups of simulations differ
slightly in other aspects, namely
\begin{itemize}
     \item[a)] Annihilation in the Earth and the Sun.  In this case we
     are mainly interested in the flux of high energy muon neutrinos
     and neutrino-induced muons at a neutrino telescope.  We simulate 6
     `fundamental' annihilation channels, $c\bar{c}$, $b\bar{b}$,
     $t\bar{t}$, $\tau^+\tau^-$, $W^+W^-$ and $Z^0 Z^{0}$ (if kinematically allowed) 
     for each 
     mass listed above. The lighter leptons and
     quarks do not contribute significantly and can safely be
     neglected. Pions and kaons get stopped
     before they decay and are thus made stable in the \code{Pythia}
     simulations so that they do not produce any neutrinos.  For
     annihilation channels containing Higgs bosons, we can calculate
     the yield from these fundamental channels by letting the Higgs
     bosons decay in flight (see below).  We also take into account
     the energy losses of $B$-mesons in the Sun and the Earth by
     following the approximate treatment of \cite{RS} but with updated
     $B$-meson interaction cross sections as given in
     \cite{joakimthesis}.  Neutrino-interactions on the
     way out of the Sun are simulated with \code{Pythia} including neutral current
     interactions and charged-current interactions as a neutrino-loss.  The
     neutrino-nucleon charged current interactions close to the
     detector are also simulated with \code{Pythia} and finally the
     multiple Coulomb scattering of the muon on its way to the detector
     is calculated using distributions from~\cite{PDG}. 
     For more details on these simulations see~\cite{Edpre,Angdist}.

     For each annihilation channel and mass we simulate $1.25 \times
     10^{6}$ annihilations and tabulate the final results as a
     neutrino-yield, neutrino-to-muon conversion rate and a muon yield
     differential in energy and angle from the center of the Sun/Earth.
     We also tabulate the integrated yield above a given threshold and
     below an open-angle $\theta$. We assumed throughout that the
     surrounding medium is water with a density of 1.0 g/cm$^3$. Hence,
     the neutrino-to-muon conversion rates have to be multiplied by the
     density of the medium. In the muon fluxes, the density cancels out
     (to within a few percent). For the neutrino-nucleon cross sections, we have
     used the parameterizations in \cite{joakimthesis}.

     \item[b)] Annihilation in the halo.  The simulations are here
     simpler since we do not have a surrounding medium that can stop the
     annihilation products.  We here simulate for 8 `fundamental'
     annihilation channels $c\bar{c}$, $b\bar{b}$,
     $t\bar{t}$, $\tau^+\tau^-$, $W^+W^-$, $Z^0 Z^{0}$, $g g$ and
     $\mu^{+} \mu^{-}$. Compared to the simulations in the Earth and
     the Sun, we now let pions and kaons decay and we also let
     antineutrons decay to antiprotons. For each mass we simulate
     $2.5 \times 10^{6}$ annihilations and tabulate the yield of
     antiprotons, positrons, gamma rays (not the gamma lines),
     muon neutrinos and neutrino-to-muon conversion rates and the
     neutrino-induced muon yield, where in the last two cases the
     neutrino-nucleon interactions has been simulated with {\sc
     Pythia} as outlined above.
\end{itemize}

With these simulations, we can calculate the yield for any of the above mentioned 
particles for a given MSSM model.  For the Higgs bosons, which decay
in flight, an integration over the angle of the decay products with
respect to the direction of the Higgs boson is performed.  Given the
branching ratios for different annihilation channels it is then
straightforward to compute the muon flux above any given energy
threshold and within any angular region around the Sun or the center
of the Earth.

\subsection{Neutrinos from the Sun and Earth}

One of the most promising indirect detection method \cite{neutrinos} relies on the
fact that scattering of halo neutralinos by the Sun and the planets,
in particular the Earth, during the several
billion years that the Solar system has existed, will have trapped
these neutralinos within these astrophysical bodies. Being trapped
within the Solar or terrestrial material, they will sink towards
the center, where a considerable enrichment and corresponding
increase of annihilation rate will occur.

Searches for neutralino annihilation into neutrinos
will be subject to  extensive experimental investigations in view
of the new neutrino telescopes (AMANDA, IceCube, Baikal, NESTOR, ANTARES)
planned or under construction \cite{halzen}. A high-energy
neutrino signal in the direction of the center of the Sun or Earth
is an excellent experimental signature which may stand up against
the background of neutrinos generated by cosmic-ray interactions in the
Earth's atmosphere.

The detector energy threshold for
``small''
neutrino telescopes like Baksan, MACRO and Super-Kamiokande is
around 1 GeV.
Large area neutrino telescopes in the ocean  or in Antarctic ice
typically
have thresholds of the order of tens of GeV, which makes them
sensitive mainly to heavy neutralinos (above 100 GeV)
\cite{begnu2}.
In \ds, the low energy cut-off can be set. 

Final states
which give a hard neutrino spectrum (such as heavy quarks, $\tau$
leptons and $W$ or $Z$ bosons) are usually more important
than the soft spectrum arising from light quarks and gluons.

Neutralinos are steadily being trapped in the Sun or Earth by
scattering, whereas annihilations take them away.
The balance between capture and annihilation
has the solution for the annihilation rate implemented in \ds\
\beq
\Gamma_A={C_c\over 2} \tanh^2\left({t\over \tau}\right),\label{eq:tanh}
\eeq
where the equilibration time scale $\tau=1/\sqrt{C_c C_a}$, with
$C_c$ being the capture rate and $C_a$ being related to the annihilation efficiency.  In most
cases for the Sun, $\tau$ is much smaller than a few billion years, and therefore
equilibrium is often a good approximation ($\dot N(t)=0$). This means
that it is the capture rate which is the important quantity that
determines the neutrino flux. For the Earth, $\tau$ is, on the
other hand, usually of the same order as, or much larger than, the age of the solar system, and equilibrium has often not occurred.
In either case, in the program we keep the exact
formula (\ref{eq:tanh}) (except for the modifications needed for a Damour-Krauss population of
WIMPs, discussed below).

For the actual capture rate calculations we have several expressions implemented in \ds. As a default, we use the full expressions given in appendix A of \cite{Gould87} where we numerically integrate over the velocity distribution. To speed-up the calculations, it is possible to perform this integration only once and use a saved tabulated version for subsequent calls.  In the capture rate calculations we also need the density profile of the Earth and the Sun and the chemical composition as a function of radius. For the Sun we use the solar model \code{BP2000} \cite{bp2000}, complemented with the estimates of the mass fractions of the heavier elements from \cite{GrevSau98}. For the Earth, we use the density profile of \cite{EncBrit} with the compositions given in \cite{earthcomp} (see \cite{le} for a table of these).
For comparison, the approximate capture rate expressions in \cite{jkg} are also available.

The capture rate induced by scalar (spin-independent) interactions
between the neutralinos and the nuclei in the interior of the Earth or
Sun is the most difficult one to compute, since it depends sensitively
on the Higgs mass, form factors, and other poorly known quantities.
However, this spin-independent capture rate calculation is the same as
for direct detection treated in Section~\ref{subs:direct}.  Therefore,
there is a strong correlation between the neutrino flux expected from
the Earth (which is mainly composed of spin-less nuclei) and the
signal predicted in direct detection experiments \cite{begnu2,kamsad}.

Due to
the low counting rates for the spin-dependent interactions in
terrestrial detectors, high-energy neutrinos from the Sun constitute a
competitive and complementary neutralino dark matter search.  Of
course, even if a neutralino is found through direct detection, it
will be extremely important to confirm its identity and investigate
its properties through indirect detection.  In particular, the mass
can be determined with reasonable accuracy by looking at the angular
distribution of the detected muons \cite{EG,BEK}.

The capture rate in the Earth is
dominated by scalar interactions, where there may be kinematic and
other enhancements, in particular if the mass of the neutralino almost
matches one of the heavy elements in the Earth. As shown by Gould \cite{Gould91}, the Earth does not dominantly capture WIMPs from the halo directly. Instead, the Earth captures WIMPs that, due to gravitational interactions, have diffused around and become bound to the solar system. However, solar depletion of these bound WIMPs could be an important effect \cite{gouldnew}, and as a default, \ds\ uses a new estimate of the velocity distribution in the solar system, where these solar depletion effects have been included \cite{le}. It is possible to change to a standard Gaussian distribution if the user prefers.

There is also a possibility that there exists a special population of WIMPs, the Damour-Krauss population \cite{dk}, arising from WIMPs that have just skimmed the Sun's surface. As an optional choice, this population can be included in the calculation \cite{dkpop}.
The enhancement caused by the new population is only important for a 
neutralino mass less than 150 - 170 GeV (the exact number depending on
details about the angular momentum distribution \cite{dkpop}).
The total capture rate  is computed according
to the formulas in \cite{dkpop}, which take into account that the annihilation
rates from the Earth will in general depend on time in a different
way than the simple result in Eq.~(\ref{eq:tanh}).

\subsection{Indirect searches through antimatter signals}

Pair annihilation of neutralinos in the Galactic halo 
produces the same amounts of matter and antimatter. As antimatter seems to 
be scarce in the Universe, apparently with no standard primary sources, 
there is a chance that by measuring antimatter in cosmic ray fluxes one may 
find evidence of the existence of dark matter particles
(see, e.g.,~\cite{silksre,SRW,kamturner,Chardonnet,bottinolast,pbarpaper,epluspaper,dbar, donato-pbar}). In the current 
release of the code we consider neutralino induced fluxes of antiprotons, 
positrons and antideuterons. To produce estimates of such fluxes, there are 
several steps one needs to follow: {\sl i)} Evaluate for each dark matter 
candidate the probability for a pair annihilation to take place locally 
in space, i.e. compute
$1/2\,(\rho_{\chi}(\vec{x}\,) / {m_{\chi}})^2 \sigma_{\rm ann}v$;\footnote{The 
origin of the factor of 1/2, missing in earlier versions
of \ds, is explained in Sect.~\protect\ref{subs:note}.} {\sl ii)} Estimate 
the production rate of a given species by folding together, for each model, 
the branching ratios for the annihilation into a given two-body final state 
with the Monte Carlo simulation of the hadronization and/or decay of that 
state, as described in Section~\ref{sec:mcsim} (except for $\bar{D}$ sources 
where we have implemented the prescription suggested in Ref.~\cite{dbar} to 
convert from the $\bar{p}$ yield); {\sl iii)} Propagate these sources
through the Galactic magnetic fields to make predictions for the induced 
cosmic ray fluxes at the Sun's location in the Galaxy; {\sl iv)} Include 
the effect of solar modulation to propagate the fluxes from the interstellar 
medium to our location inside the solar system. The implementation in \ds\ 
is written in a modular format which allows the user to eventually modify 
and/or replace each of the four steps above independently.
 
At tree level, the relevant final states of neutralino pair annihilations
for $\bar{p}$ and $\bar{D}$ 
production are $q \bar{q}$, $W^{+} W^{-}$, $Z^{0} Z^{0}$,
$W^{+} H^{-}$, $Z H_{1}^{0}$, $Z H_{2}^{0}$, $H_{1}^{0} H_{3}^{0}$ and
$H_{2}^{0} H_{3}^{0}$; among the $q \bar{q}$ states we have included in \ds\ 
all the heavier quarks ($c$, $b$ and $t$). In addition, we have 
included the $Z \gamma$ (\cite{ub}) and the 2 gluon (\cite{bua}; \cite{lp}) 
final states which occur at one loop-level. The same list of final states
is implemented for positrons, with the addition of $\ell^+ \ell^-$.
The $e^+ e^-$ final state gives rise to a positron monochromatic
source (however this is negligible in most particle physics setups, and
tends to be smeared out by propagation).

To model the propagation we have considered a semi-empirical diffusion 
equation~\cite{Berezinskii,Gaisserbook}, in the steady state approximation, 
applied to a two-dimensional model with cylindrical symmetry and with free 
escape boundary conditions. The parameters in such a model should be fixed in 
agreement with values inferred from available cosmic ray data in analogous 
propagation models (see, e.g., the {\sc Galprop} package~\cite{galprop}). For 
simplicity, rather than considering the most general setup, we restrain 
to cases in which, still including all  relevant effects, the Green 
function of the transport equation can be derived analytically, so that 
we can avoid the CPU time-consuming problem of having to solve a partial 
differential equation for each dark matter candidate. Therefore, we do not 
include reacceleration effects but mimic them through a diffusion 
coefficient which has a broken power law in rigidity as functional form. 
Also , for antiprotons~\cite{pbarpaper} and antideuterons we neglect energy losses 
(whenever 
a scattering with a nucleus takes place the particle is removed), while 
for positrons~\cite{epluspaper} we consider an average over space of the energy loss effect 
due to inverse Compton on starlight and the cosmic microwave background, and in
addition the synchrotron radiation from the effects of the galactic magnetic
field. 
For comparison, we allow also the option to treat the propagation of 
antiprotons according to the propagation models by Chardonnet et al.\ 
\cite{Chardonnet} and Bottino et al.\ \cite{bottinolast}, while for the 
positron we have implemented the option to use the leaky-box treatment 
given by Kamionkowski and Turner \cite{kamturner} or the numerical Green 
functions derived by Moskalenko and Strong \cite{MoskStrong98}
with the {\sc Galprop} code~\cite{galprop} (the latter two cases however 
cannot be interfaced to a generic axisymmetric dark matter density 
profile, as for our default propagation model).

Given a set of parameters for the propagation model, and a given neutralino 
number density profile or a given probability distribution of small clumps, 
the effect of propagation, in the cases we have considered, can be 
factorized out into effective energy-dependent ``diffusion times'', 
$\tau_{\bar{p}}(T)$,  $\tau_{\bar{D}}(T)$ and  $\tau_{e^+}(T)$, 
which are independent 
of the parameters defining the particle physics setup for the dark 
matter candidate. In some cases, such as when considering very large 
samples of neutralino candidates or when implementing singular halo 
profiles for which the computation of the diffusion times can be very 
CPU time-consuming, it is advantageous to exploit the option provided 
by \ds\ to tabulate the diffusion times over a given energy range 
(optionally saving the tabulation to disk for later use) and use an 
interpolation between tabulated values, rather than linking directly 
to the computation for each dark matter candidate. 

Finally, regarding solar modulation, we implement the one parameter 
model based on the analytical force-field approximation by Gleeson \& 
Axford \cite{GleesonAxford} for a spherically symmetric model. This 
approach is expected to be slightly less accurate than, e.g., the full
numerical solution of the propagation equation of the spherically symmetric 
model~(\cite{fisk}), but again it is much less CPU time-consuming. \ds\ 
allows an output with both solar modulated and local interstellar fluxes, 
and the latter can eventually be solar modulated by the user with more 
sophisticated methods. For the positrons we allow for the option to 
reduce the effects of solar modulation by considering the positron 
fraction, $e^+/(e^+ + e^-)$, rather than the absolute positron fluxes. 
In this case an estimate of the background is needed: in \ds\ we provide
background $e^+$ and $e^-$ fluxes taken from~\cite{MoskStrong98}.

\subsection{Gamma rays}

Gamma-rays have a low 
enough cross section on gas and dust that the Galaxy is 
essentially transparent to them (except perhaps in the innermost part, 
very close to the region where a massive black hole is inferred); 
absorption by starlight and infrared background becomes efficient
only for very far away sources and high-energy photons.

The bulk of the gamma-rays from neutralino annihilations arise in the
decay of neutral pions produced in the fragmentation processes initiated
by tree level final states~\cite{oldcontga,lpj,gahalo}
(analogously to the other halo signals,
in \ds\ we include all tree level final states and make use of a
Monte Carlo simulation for fragmentation and decay processes, see 
Section~\ref{sec:mcsim}). However, $\pi^0$ 
production is common also to other astrophysical processes, and this
may turn out to be a limiting factor to disentangle dark matter sources.
At the same time, however, a relevant gamma-ray contribution may arise 
directly (at one-loop level) in two body final states; although such 
photons are much fewer than those from $\pi^0$ decays they have a much 
better signature: neutralinos annihilating in the galactic halos move 
with a velocity of the order $v/c \sim$ 10$^{-3}$, hence these outgoing 
photons (as any particle in any of the allowed two body final states)
will then be nearly monochromatic, with energy of the order of the
neutralino mass~\cite{charm,oldlines,jkline,lp,ub,lpj}. 
There is no other known astrophysical source with such a signature:
the detection of a line signal out of a spectrally smooth gamma-ray 
background would be a spectacular confirmation of the existence of dark 
matter in form of exotic massive particles.

\subsubsection{Sources and fluxes}

Following the discussion in \cite{lpj}, the monochromatic gamma-ray flux measured in a detector with angular acceptance $\Delta\Omega$ is:
\begin{equation}
\Phi_{\gamma}(\psi, \Delta\Omega) = 0.94 \cdot 10^{-11}
\left(\frac{N_{\gamma}\;v\sigma_{X^0\gamma}}
{10^{-29}\ {\rm cm}^3 {\rm s}^{-1}}\right)
\left( \frac{10\,\rm{GeV}}{M_\chi}\right)^2 
\langle\,J\left(\psi\right)\rangle_{\Delta\Omega}\times\Delta\Omega 
\;\rm{cm}^{-2}\;\rm{s}^{-1}\; \rm{sr}^{-1}\; ,
\label{signal}
\end{equation}
where $\psi$ is an angle to label the direction of observation and where 
$N_{\gamma} = 2$ for $\chi\,\chi\rightarrow \gamma\,\gamma$, 
$N_{\gamma} = 1$ for $\chi\,\chi\rightarrow Z\,\gamma$.
Here the dimensionless line-of-sight dependent function is 
\beq
J\left(\psi\right) = \frac{1} {8.5\, \rm{kpc}}
\cdot \left(\frac{1}{0.3\,{\rm GeV}/{\rm cm}^3}\right)^2
\int_{line\;of\;sight}\rho_{\chi}^2(l)\; d\,l(\psi)\;,
\label{eq:jpsi}
\eeq
and its angular average over the resolution solid angle $\Delta\Omega$ is
\begin{equation}
\langle\,J\left(\psi\right)\rangle_{\Delta\Omega}
= \frac{1}{\Delta\Omega} \int_{\Delta\Omega} d\Omega^{\prime}
J\left(\psi^{\prime}\right)\;, 
\label{eq:jave}
\end{equation}

Analogously,
the gamma-ray flux with continuum energy spectrum is obtained by replacing
$N_{\gamma}\;v\sigma_{X^0\gamma}$ with 
$\sum_f dN_{\gamma}^f/dE\;v\sigma_{f}$, where the sum is over all tree
level final states. 

Finally, the formalism we introduced can be used also to estimate
the flux in the simple case of a single source which, for a given 
detector, can be approximated as point-like. 
If such a
source is in the direction $\psi$ at a distance $d$, Eq.~(\ref{eq:jave})
becomes:
\begin{equation}
\langle\,J\left(\psi\right)\rangle_{\Delta\Omega}
= \frac{1} {8.5\, \rm{kpc}}
\cdot \left(\frac{1}{0.3\,{\rm GeV}/{\rm cm}^3}\right)^2
\cdot \frac{1}{d^2} \cdot \frac{1}{\Delta\Omega}
\int d^3r \;\rho_{\chi}^2(\vec{r}) 
\end{equation}
where the integral is over the extension of the source (much smaller
than $d$).

Several targets have been discussed as sources of gamma-rays from the
annihilation of dark matter particles. An obvious source is the dark
halo of our own galaxy~\cite{galga} and in particular the Galactic center,
as the dark matter density profile is expected, in most models, to be 
peaked towards it, possibly with huge enhancements close to the central 
black hole. The Galactic center is an ideal target for both ground-
and space-based gamma-ray telescopes. As satellite experiments have 
much wider field of views and will provide a full sky coverage,
they will test the hypothesis of gamma-rays emitted in clumps of dark 
matter which may be present in the 
halo~\cite{clumpyga,gahalo,clumpy,clumpybeg}. 
Another possibility which has been considered is the case of 
gamma-ray fluxes from external nearby galaxies~\cite{extergal}. 
Furthermore, it has 
been proposed to search for an extragalactic flux originated by all 
cosmological annihilations of dark matter 
particles~\cite{extragal,extragalbeu}.

\ds\ is suitable to compute the gamma-ray flux from all these (and possibly 
other) sources. Two cases are fully included in the package:
assuming that neutralinos are smoothly distributed in the Galactic halo
with $\rho_{\chi}$ equal to the dark matter density profile, in 
\ds\ Eq.~(\ref{eq:jave}) is computed for a specified halo profile and 
any given $\psi$ and $\Delta\Omega$~\cite{lpj}. 
The second option deals with the
case of a portion of dark matter being in the form of clumps, each of
which is treated as a non-resolvable source in the detector, distributed
in the Galaxy according to a probability distribution which
can be specified by the user; in \ds\ the default probability distribution stems
from the assumption that it follows the dark matter density profile (see \cite{clumpy}
for details). It is straightforward to extend this to all other 
astrophysical sources; in case of cosmological sources one has just 
to pay attention to include redshift effects and absorption on starlight 
and infrared background, see~\cite{extragalbeu}.

The case of the possible enhancement, a ``spike'' in the vicinity of
the galactic center \cite{gs} should be kept in mind. However,
since there is no consensus in the literature \cite{paolospike,bertone,aloisio}
about important quantities
for the calculation such as the magnetic field radial profile and the
optical depth for synchrotron self-absorption, we have chosen not to 
include routines for the effects of this possible enhancement
of gamma rays and neutrinos. And the very existence of a spike is dependent
on fine details, still unknown for the Milky Way \cite{kamullio,merritt}.


\subsubsection{Gamma rays with continuum energy spectrum}
The gamma-ray flux produced in neutralino annihilations through 
$\pi^0$ decays can be large but in general lacks distinctive features.
The photon spectrum in the process of a pion decaying into $2\gamma$ is, 
independent of the pion energy, peaked at half of the $\pi^0$ mass, 
about 70~MeV, and symmetric with respect to this peak if plotted in 
logarithmic variables (i.e.\ $dN_\gamma/dE$ vs.\ $\log E$). 
Of course, this is true both for pions produced in 
neutralino annihilations and, e.g., for those generated by cosmic ray 
protons interacting with the interstellar medium.

When considered together with the cosmic ray induced Galactic gamma-ray
background, the neutralino induced signal looks like a component analogous
to the secondary flux due to nucleon nucleon interactions: it is
dwarfed by the bremsstrahlung component at low energy, while it may 
be the dominant contribution at energies above 1 GeV or so. 
In \ds\ the continuum gamma flux from all annihilation channels is computed
and may be easily obtained for a given energy or energy threshold.

\subsubsection{$\chi\chi\to \gamma\gamma$}

At the one-loop level, it is possible to get two-body final
states containing one or two photons, with  a distinctive
signature which may provide a ``smoking gun'' for dark matter.
The drawback is of course that the processes are loop-suppressed,
so one probably needs a halo with a large central concentration,
or small-scale structure (``clumps'' of dark matter) to detect
a signal.

In \ds\ the full expression for the annihilation cross section of
the process
\beq
\tilde{\chi}^{0}_{1} + \tilde{\chi}^{0}_{1} \rightarrow \gamma
+\gamma
\eeq
is computed at full one loop level, in the limit of vanishing relative 
velocity of the neutralino pair, i.e. the case of interest for neutralinos
in galactic halos; the outgoing photons have an energy equal to 
the mass of ${\chi}^{0}_{1}$:
\beq
E_{\gamma} = m_{\chi}.
\eeq

The total amplitude is implemented in \ds\ as the sum of the contributions 
obtained from four different classes of diagrams:
\begin{eqnarray*}
\tilde{\mathcal{A}}=\tilde{\mathcal{A}}_{f\tilde{f}}+
  \tilde{\mathcal{A}}_{H^+}+\tilde{\mathcal{A}}_{W}+\tilde{\mathcal{A}}_{G},
\end{eqnarray*}
where the indices label the particles in the internal loops, i.e.,
respectively, fermions and sfermions, charged Higgs and charginos, 
W-bosons and charginos, and, in the gauge we chose, charginos and Goldstone 
bosons. For every $\mathcal{A}$ term, real
and imaginary parts are separately computed; 
the full set of analytic formulas are 
given in \cite{lp}, following the notation of \cite{jkline}, where some of 
the contributions were first computed. They are rather lengthy expressions
with non trivial dependences on various combinations of parameters in 
the MSSM. 

The branching ratio for neutralino annihilations into $2\gamma$ is 
typically not larger than 1\%, with the largest values of
$v\sigma_{2\gamma}$, for neutralinos with a cosmologically significant
relic abundance, in the range $10^{-29}$--$10^{-28}$~cm$^3$s$^{-1}$.
Such values may be large enough for the discovery of this signal
in upcoming measurements; at the same time it should be kept in mind
that very low values for the cross section are possible as well.

In the very high mass range above a TeV, it has been suggested that the line
rates may be very much larger due to binding effects and resonant conversion
between neutralinos and charginos \cite{nojiri03}. In the present version
of \ds\ we have not included these effects, however.

\subsubsection{$\chi\chi\to Z\gamma$}

The process of neutralino annihilation into a photon and a Z$^0$ boson
\cite{ub}
\beq
\tilde{\chi}^{0}_{1} + \tilde{\chi}^{0}_{1} \rightarrow \gamma
+ Z^0
\eeq
also gives a nearly
monochromatic line (with a small smearing caused by the finite
  width of the $Z^0$ boson, in any case unlikely to be important for current or
  proposed gamma ray experiments), with  energy
\beq
E_{\gamma} = M_{\chi} - \frac{m_Z^2}{4\,M_{\chi}}.
\eeq

The steps followed in \ds\
to compute the cross section are essentially the same
as those described for the  $2\gamma$ case.
Again the total amplitude is obtained by summing the contribution
from four classes of diagrams and by splitting for each of them 
real and imaginary parts. The analytic formulas were derived in 
\cite{ub}, and are much less compact than those 
obtained for the process of neutralino annihilation into two photons.

The maximum value of $v\sigma_{Z\gamma}$, for neutralinos with a 
cosmologically significant relic abundance, is around 
$2\cdot 10^{-28}$~cm$^3$s$^{-1}$ and occurs for nearly
pure higgsinos. In the heavy mass range, the value of $v\sigma_{Z\gamma}$
reaches a plateau of around $0.6 \cdot 10^{-28}$~cm$^3$s$^{-1}$. This
interesting effect of a non-diminishing cross section with Higgsino mass
(which is due to a contribution to the real part of the amplitude)
is also valid for the $2\gamma$ final state in the corresponding limit, 
with a value of $1\cdot 10^{-28}$~cm$^3$s$^{-1}$ \cite{lp}.
Since the gamma-ray background drops rapidly with increasing photon
energy, these processes may be interesting for detecting dark
matter neutralinos near the upper range of allowed neutralino masses.

Whenever the lightest neutralino contains a significant Wino or Higgsino
component the value of $v\sigma_{Z\gamma}$ may be as large as, or even larger 
than, twice the value of $v\sigma_{2\gamma}$. It is therefore usually
not a good approximation to neglect the $Z\gamma$ state compared to
$2\gamma$.

\subsection{Neutrinos from the halo}

Usually, the flux of neutrinos from annihilation of neutralinos in
the Milky Way halo is too small to be detectable, but for some clumpy
or cuspy models, or for annihilation in a possible spike around the central black hole, 
it might be detectable. The calculation of the
neutrino-flux follows closely the calculation of the continuous gamma
ray flux, with the main addition that neutrino interactions close to
the detector are also included. Hence, both the neutrino flux and the
neutrino-induced muon flux can be obtained. The neutrino to muon
conversion rate in the Earth can also be obtained.
The neutrino rate from other sources than the interior of the Earth or the Sun
is generally negligible. If there would exist a spike at the galactic center
\cite{gs}, there may in some cases be a detectable flux. These neutrino rates
are constrained by existing limits on the gamma-ray flux \cite{gs,bertone2}.

\section{Examples of results obtained with \ds}

We will here go through a set of benchmark models as examples of the performance of \ds. We will consider two popular setups.
We will start with a set of mSUGRA models and then turn to more general MSSM models.
We will here use the default \ds\ setup, in particularÊan isothermal sphere with a Maxwell-Boltzmann velocity distribution for the halo model.

\subsection{Benchmark models in mSUGRA}

We will consider here some of the benchmark models from Battaglia et al.\ \cite{bat-bench}. In table \ref{tab:bat-bench} we list the properties of the selected models, as derived by \ds\ using to the \code{ISASUGRA 7.69} package to describe the renormalization group running of the theory from the grand unification scale to the low energy scale.  The models we are focusing on are those with a top mass of 175 GeV and that are still viable with \code{ISASUGRA 7.69}. As the table in \cite{bat-bench} was produced with \code{ISASUGRA 7.67}, some differences occur due to different versions of the codes, e.g. model M is no longer physical in \code{ISASUGRA 7.69}, and thus not included here.

\begin{sidewaystable}
  \tiny
  \centering 
  \begin{tabular}{ccccccccccc}
 Model & A' & B' & C' & D' & G' & H' & I' & J' & K' & L' \\ \hline
$m_{1/2}$ [GeV] & 600 & 250 & 400 & 525 & 375 & 935 & 350 & 750 & 1300 & 450 \\
$m_0$ [GeV] & 107 & 57 & 80 & 101 & 113 & 244 & 181 & 299 & 1001 & 303 \\
$\tan \beta$ & 5 & 10 & 10 & 10 & 20 & 20 & 35 & 35 & 46 & 47 \\
${\rm sign}(\mu)$ & + & + & + & - & + & + & + & + & - & + \\ \hline \hline
 $m_\chi$ [GeV]  & 242.8 &  94.9 & 158.1 & 212.4 & 148.0 & 388.4 & 138.1 & 309.1 & 554.2 & 181.0 \\ \hline
 $B(b\to s \gamma)\times10^4$ & $3.96$ & $3.88$
 & $3.95$ & $4.39$ & $3.67$ 
 & $3.84$ & $3.33$ & $3.84$ 
 & $4.40$ & $3.62$ \\
 $a_\mu \times 10^{10}$ & 
 $1.3$ &  $18$ &  $6.7$ &
 $-4.4$ &  $16$ &  $2.7$ &
 $33$ &  $7.4$ &  $-3.3$ &
  $26$ \\
 Incl. coanns & $\tilde{e}_2$ $\tilde{\mu}_2$ $\tilde{\tau}_1$ $\chi_1^0$ 
     & $\tilde{e}_2$ $\tilde{\mu}_2$ $\tilde{\tau}_1$ $\chi_1^0$ 
     & $\tilde{e}_2$ $\tilde{\mu}_2$ $\tilde{\tau}_1$ $\chi_1^0$ 
     & $\tilde{e}_2$ $\tilde{\mu}_2$ $\tilde{\tau}_1$ $\chi_1^0$ 
     & $\tilde{e}_2$ $\tilde{\mu}_2$ $\tilde{\tau}_1$ $\chi_1^0$ 
     & $\tilde{e}_2$ $\tilde{\mu}_2$ $\tilde{\tau}_1$ $\chi_1^0$ 
     & $\tilde{\tau}_1$ $\chi_1^0$ 
     & $\tilde{e}_2$ $\tilde{\mu}_2$ $\tilde{\tau}_1$ $\chi_1^0$ 
     & $\chi_1^0$ 
     & $\tilde{\tau}_1$ $\chi_1^0$ \\
 $\Omega_\chi h^2$  & 0.0929 & 0.1213 & 0.1149 & 0.0864 & 0.1294 & 0.1629 & 0.1319 & 0.0966 & 0.0863 & 0.0988 \\
 $\Phi_\mu^\oplus$ best [km$^{-2}$ yr$^{-1}$] & $2.46 \cdot 10^{-10}$ & $1.50 \cdot 10^{-5}$
 & $9.76 \cdot 10^{-9}$ & $3.76 \cdot 10^{-14}$ 
 & $3.36 \cdot 10^{-7}$ & $1.51 \cdot 10^{-12}$ 
 & $1.89 \cdot 10^{-5}$ & $2.59 \cdot 10^{-10}$ 
 & $4.59 \cdot 10^{-16}$ & $1.07 \cdot 10^{-5}$ \\
 $\Phi_\mu^\oplus$ gauss [km$^{-2}$ yr$^{-1}$] & $3.03 \cdot 10^{-9}$ & $5.18 \cdot 10^{-5}$
 & $7.15 \cdot 10^{-8}$ & $3.92 \cdot 10^{-13}$ 
 & $2.27 \cdot 10^{-6}$ & $3.50 \cdot 10^{-11}$ 
 & $1.17 \cdot 10^{-4}$ & $4.40 \cdot 10^{-9}$ 
 & $1.48 \cdot 10^{-14}$ & $9.20 \cdot 10^{-5}$ \\
 $\Phi_\mu^\odot$ [km$^{-2}$ yr$^{-1}$] & $1.22 \cdot 10^{-2}$ & $1.47 \cdot 10^{1}$
 & $5.50 \cdot 10^{-1}$ & $3.01 \cdot 10^{-2}$ 
 & $3.33 \cdot 10^{0}$ & $1.76 \cdot 10^{-4}$ 
 & $4.45 \cdot 10^{0}$ & $1.44 \cdot 10^{-2}$ 
 & $3.42 \cdot 10^{-3}$ & $2.32 \cdot 10^{0}$ \\
 $\sigma_p^{SI}$ std [pb] & $3.02 \cdot 10^{-10}$ & $5.69 \cdot 10^{-9}$
 & $8.22 \cdot 10^{-10}$ & $2.39 \cdot 10^{-12}$ 
 & $1.98 \cdot 10^{-9}$ & $8.38 \cdot 10^{-11}$ 
 & $7.93 \cdot 10^{-9}$ & $2.81 \cdot 10^{-10}$ 
 & $6.80 \cdot 10^{-14}$ & $5.94 \cdot 10^{-9}$ \\
 $\sigma_p^{SD}$ std [pb] & $2.13 \cdot 10^{-7}$ & $1.06 \cdot 10^{-5}$
 & $1.55 \cdot 10^{-6}$ & $5.02 \cdot 10^{-7}$ 
 & $2.33 \cdot 10^{-6}$ & $8.64 \cdot 10^{-8}$ 
 & $3.31 \cdot 10^{-6}$ & $2.06 \cdot 10^{-7}$ 
 & $3.64 \cdot 10^{-8}$ & $1.37 \cdot 10^{-6}$ \\
 $\sigma_p^{SI}$ dn [pb] & $3.02 \cdot 10^{-10}$ & $5.71 \cdot 10^{-9}$
 & $8.23 \cdot 10^{-10}$ & $2.63 \cdot 10^{-12}$ 
 & $1.99 \cdot 10^{-9}$ & $8.40 \cdot 10^{-11}$ 
 & $7.99 \cdot 10^{-9}$ & $2.83 \cdot 10^{-10}$ 
 & $7.95 \cdot 10^{-14}$ & $5.97 \cdot 10^{-9}$ \\
 $\sigma_p^{SD}$ dn [pb] & $2.09 \cdot 10^{-7}$ & $1.05 \cdot 10^{-5}$
 & $1.53 \cdot 10^{-6}$ & $4.93 \cdot 10^{-7}$ 
 & $2.30 \cdot 10^{-6}$ & $8.51 \cdot 10^{-8}$ 
 & $3.27 \cdot 10^{-6}$ & $2.04 \cdot 10^{-7}$ 
 & $3.60 \cdot 10^{-8}$ & $1.36 \cdot 10^{-6}$ \\
$N_{\gamma \mbox{~cont.}}(\sigma v)$ [$10^{-29}$ cm$^3$ s$^{-1}$] 
& 120 & 782 & 195 & 63.6 & 1032 & 86.5 & 6303 & 930 & 70803 & 18739 \\
$2(\sigma v)_{\gamma \gamma}$ [$10^{-29}$ cm$^3$ s$^{-1}$] 
& 1.0 & 2.6 & 1.7 & 1.3 & 1.4 & 0.35 & 0.77 & 0.33 & 0.017 & 0.39 \\
$(\sigma v)_{Z \gamma}$ [$10^{-29}$ cm$^3$ s$^{-1}$] 
& 0.17 & 0.31 & 0.26 & 0.19 & 0.18 & 0.051 & 0.083 & 0.040 & 0.0022 & 0.037 \\
$\Phi_{e^+}$ [GeV$^{-1}$ cm$^{-2}$ s$^{-1}$ sr$^{-1}$] 
& $1.5 \cdot 10^{-11}$ & $5.4 \cdot 10^{-10}$ & $7.0 \cdot 10^{-11}$
& $2.2 \cdot 10^{-11}$ & $4.1 \cdot 10^{-10}$ & $7.7 \cdot 10^{-12}$
& $2.3 \cdot 10^{-9}$  & $1.1 \cdot 10^{-10}$ & $3.7 \cdot 10^{-9}$
& $4.7 \cdot 10^{-9}$ \\
$\Phi_{\bar{p}}$ [GeV$^{-1}$ cm$^{-2}$ s$^{-1}$ sr$^{-1}$] 
& $6.0 \cdot 10^{-11}$ & $2.2 \cdot 10^{-9}$  & $1.5 \cdot 10^{-10}$
& $2.1 \cdot 10^{-11}$ & $9.0 \cdot 10^{-10}$ & $7.2 \cdot 10^{-12}$
& $6.7 \cdot 10^{-9}$  & $1.4 \cdot 10^{-10}$ & $2.5 \cdot 10^{-9}$
& $1.0 \cdot 10^{-8}$ \\ 
$\Phi_{\bar{D}}$ [GeV$^{-1}$ cm$^{-2}$ s$^{-1}$ sr$^{-1}$] 
& $2.3 \cdot 10^{-15}$ & $9.1 \cdot 10^{-14}$ & $5.8 \cdot 10^{-15}$ 
& $8.7 \cdot 10^{-16}$ & $3.6 \cdot 10^{-14}$ & $2.7 \cdot 10^{-16}$
& $2.6 \cdot 10^{-13}$  & $5.1 \cdot 10^{-15}$ & $9.0 \cdot 10^{-14}$
& $4.0 \cdot 10^{-13}$ \\ \hline
\end{tabular}
  \caption{Relic density and various rates for the benchmark models of \cite{bat-bench}. There are five free parameters in mSUGRA: $\tan\beta$, sign($\mu$), $m_{1/2}$, $m_0$ and $A_0$. The latter three are the unification values (at the grand unification scale) of the soft supersymmetry breaking fermionic mass parameters, scalar mass parameters and trilinear scalar coupling parameters,  respectively. All the benchmark models have $A_0 = 0$. We have here used \code{ISASUGRA 7.69} for the RGE-running (but have not taken the b, t, and $\tau$ Yukawa couplings from \code{ISASUGRA}). The neutrino-induced muon fluxes from the Earth and the Sun are for a threshold of 1 GeV\@. `best' refers to the suppressed fluxes resulting from the estimate in \protect\cite{le} and `gauss' refers to the usual approximation of a Gaussian velocity distribution of neutralinos for capture in the Earth. The scattering cross sections are given here evaluated both with the standard expressions (labelled `std') and with the Drees and Nojiri expressions \protect\cite{dn} (labelled `dn'). The continuum $\gamma$'s are given in terms of the number of photons times the cross section, $N_{\gamma \mbox{~cont.}}(\sigma v)$, and refers to $\gamma$'s above 1 GeV\@.
The $e^+$ flux is the average solar modulated flux in the energy range 6.0--8.9 GeV, i.e.\ in one of the HEAT 94-95's bins\protect\cite{heat}. The $\bar{p}$ flux is the average solar modulated flux in the energy range 0.56--0.78 GeV, i.e.\ in one of the BESS 98 bins \protect\cite{bess}. The $\bar{D}$ flux is the average flux in the energy range 0.1--0.4 GeV, as applicable to e.g.\ the proposed GAPS probe \protect\cite{GAPSproposal}. In this case, the flux is the average of the solar modulated fluxes at solar minimum and maximum. Parameters other than those mentioned above are set to their default values as described in the text.}\label{tab:bat-bench}
\end{sidewaystable}

Our results for the relic density and for the SUSY contributions to $B(b\to s \gamma)$ and to the anomalous magnetic moment of the muon $a_\mu$, agree reasonably well with those in~\cite{bat-bench}. The expected sensitivities of future neutrino telescopes are of the order of 20 events km$^{-2}$ yr$^{-1}$ for the Earth and 50--1000 events km$^{-2}$ yr$^{-1}$ for the Sun; we see in the table that all benchmark models, unfortunately,  produce much lower fluxes. The outreach of future direct detection experiments is expected to be of the order of $10^{-9}$ pb for the spin-independent scattering cross section; we see that some of models we are considering are potentially detectable. The cross section for annihilation into gamma rays (times the number of photons produced) are also given for these models; the detectability depends strongly on the halo profile, but one can in general say that these cross sections are too low to be seen in current data, unless the halo profile is very cuspy towards the galactic center. The $e^+$ fluxes are here given in  the energy bin 6.0--8.9 GeV of the HEAT 94+95 \cite{heat} experiment. The measured $e^+$ flux is $(7.2 \pm 1.2) \times 10^{-6}$ GeV$^{-1}$ cm$^{-2}$ s$^{-1}$ sr$^{-1}$, i.e.\ the predicted fluxes are much lower than the measured one. For the antiprotons, we choose, as an example, to show the predicted (average) flux in the energy range 0.56--0.78 GeV, which is one of the BESS 98~\cite{bess} bins. The measured flux in this bin is $(1.23^{+0.38}_{-0.33}) \times10^{-6}$ GeV$^{-1}$ cm$^{-2}$ s$^{-1}$ sr$^{-1}$, i.e.\ the measured flux is much higher than the predicted one in this energy range. For the antideuterons, we show the expected flux in the energy range 0.1--0.4 GeV, which is reasonable for the proposed GAPS probe \cite{GAPSproposal}. For the antideuterons, there is essentially no background and the sensitivity is thus given by the ability to detect one antideuteron. For GAPS this corresponds to a sensitivity of about 
$2.6 \times 10^{-13}$ GeV$^{-1}$ cm$^{-2}$ s$^{-1}$ sr$^{-1}$. Hence, two of these benchmark models have high enough fluxes for being just about detectable in this way. For further examples of rates in mSUGRA, as calculated with \ds\ , see e.g.\ \cite{esu-rates}.
     

\subsection{Benchmark models in MSSM}

We will now turn to models in the MSSM framework as generated by fixing free parameters at the weak scale. We refer to the setup with seven free parameters,
i.e.  $\mu$, $M_2$, $m_A$, $\tan\beta$, $m_0$, $A_t$ and $A_b$, we described in Section \ref{sec:MSSMdef}. We will here show an example of results that can be obtained with a simple scan over this parameter space. We have generated 5000 models  assuming: $\mu \in [-3000,3000]$ GeV, $M_2 \in [-3000,3000]$ GeV, $m_A \in [100,1000]$ GeV, $\tan \beta \in [1,55]$, $m_0 \in [100,5000]$ GeV, $A_t \in [-3,3]m_0$ and $A_b \in [-3,3]m_0$.  We have then selected a few sample models with a relic density in the range $0.09 \le \Omega_\chi h^2 \le 0.11$ and with reasonably high detection rates; these are the first ten models in Table~\ref{tab:mssm}. In addition to this scan, we have also made a small scan of 300 models in which we required the mass of the $CP$-odd Higgs boson, $A$, to be in the range $m_A \in [90,150]$ GeV; model number 11 in Table \ref{tab:mssm} has been chosen from this latter scan to illustrate that higher-rate models are possible to find with these kind of dedicated scans.

\begin{sidewaystable}
  \tiny
  \centering 
  \begin{tabular}{ccccccccccc}
 Model & 1 & 2 & 3 & 4 & 5 & 6 & 7 & 8 & 9 & 10 \\ \hline
$\mu$ [GeV] & $441.9$ & $-203.0$ & $218.9$ & $153.8$ & $996.6$ & $-109.8$ 
  & $122.2$ & $-1020.2$ & $512.2$ & $-336.1$ \\ 
$M_2$ [GeV] & $-785.4$ & $329.0$ & $-361.0$ & $-213.3$ & $2446.9$ & $201.6$ 
  & $282.0$ & $2895.0$ & $776.2$ & $-438.4$ \\
$m_A$ [GeV] & 925.9 & 507.8 & 950.7 & 703.4 & 436.5 & 856.9 & 587.9 & 753.2 & 636.0 & 104.4 \\
$\tan \beta$ & 9.7 & 8.3 & 15.7 & 38.4 & 13.1 & 13.4 & 3.3 & 14.7 & 3.0 & 27.7 \\
$m_0$ [GeV] & 2675.2 & 4793.8 & 4078.7 & 4249.6 & 1573.0 & 4511.5 & 1404.5 & 4001.4 & 1084.7 & 589.8 \\
$A_t/m_0$ & $1.14$ & $-0.75$ & $2.39$ & $-1.10$ & $-2.43$ & $0.63$ & $2.32$ 
  & $-0.58$ & $2.46$ & $2.50$ \\
$A_b/m_0$ & $-1.92$ & $2.90$ & $-1.02$ & $-1.62$ & $-0.77$ & $0.47$ & $-1.81$
  & $0.62$ & $1.79$ & $0.98$ \\ \hline \hline
 $m_\chi$ [GeV]  & 382.5 & 151.7 & 165.2 & 92.2 & 98.8 & 73.2 & 78.0 & 1017.8 
  & 378.4 & 212.9 \\ 
 $Z_g$   & 0.788 & 0.690 & 0.675 & 0.672 & 0.020 & 0.377 & 0.298 & 0.005 
  & 0.907 & 0.917 \\ \hline
 $B(b\to s \gamma)\times10^4$ & $4.01$ & $4.45$ & $3.97$ & $4.37$ & $4.22$ & $4.12$ 
   & $4.42$ & $4.20$ & $4.51$ & $2.82$ \\
 $a_\mu \times 10^{10}$ & 
 $-0.63$ &  $-0.27$ &  $-0.59$ &
 $-1.26$ & $0.61$ &  $-0.45$ &
 $0.28$ &  $-0.24$ &
  $0.50$ & $15.87$ \\
 Incl. coanns & $\chi_2^+$ $\chi_{1,2,3}^0$ 
  & $\chi_2^+$ $\chi_{1,2,3}^0$ 
  & $\chi_2^+$ $\chi_{1,2,3}^0$ 
  & $\chi_2^+$ $\chi_{1}^0$ 
  & $\tilde{t_1}$ $\chi_2^+$ $\chi_{1,2,3}^0$ 
  & $\chi_2^+$ $\chi_{1}^0$ 
  & $\chi_2^+$ $\chi_{1}^0$ 
  & $\chi_2^+$ $\chi_{1,2,3}^0$ 
  & $\chi_2^+$ $\chi_{1,2,3}^0$ 
  & $\chi_2^+$ $\chi_{1,2}^0$ 
   \\
 $\Omega_\chi h^2$  & 0.0926 & 0.0982 & 0.0960 & 0.0988 & 0.0918 & 0.1000 & 0.0963
    & 0.1032 & 0.0977 & 0.0968 \\
 $\Phi_\mu^\oplus$ best [km$^{-2}$ yr$^{-1}$]
  & $7.62 \cdot 10^{-6}$ & $1.94 \cdot 10^{-5}$ & $1.05 \cdot 10^{-4}$
  & $5.63 \cdot 10^{-5}$ & $7.60 \cdot 10^{-6}$ & $3.30 \cdot 10^{-5}$
  & $3.66 \cdot 10^{-3}$ & $1.62 \cdot 10^{-8}$
  & $4.67 \cdot 10^{-5}$ & $2.17 \cdot 10^{1}$
 \\
 $\Phi_\mu^\oplus$ gauss [km$^{-2}$ yr$^{-1}$]
  & $1.62 \cdot 10^{-4}$ & $1.27 \cdot 10^{-4}$ & $7.56 \cdot 10^{-4}$
  & $1.72 \cdot 10^{-4}$ & $2.45 \cdot 10^{-4}$ & $5.58 \cdot 10^{-5}$
  & $7.43 \cdot 10^{-3}$ & $5.26 \cdot 10^{-7}$
  & $9.79 \cdot 10^{-4}$ & $2.04 \cdot 10^{2}$ \\
 $\Phi_\mu^\odot$ [km$^{-2}$ yr$^{-1}$] 
  & $6.06 \cdot 10^{1}$ & $1.56 \cdot 10^{3}$ & $1.29 \cdot 10^{3}$
  & $3.23 \cdot 10^{3}$ & $3.69 \cdot 10^{0}$ & $1.27 \cdot 10^{3}$
  & $6.77 \cdot 10^{2}$ & $3.08 \cdot 10^{-1}$
  & $1.74 \cdot 10^{1}$ & $3.66 \cdot 10^{3}$ \\
 $\sigma_p^{SI}$ std [pb] 
  & $4.08 \cdot 10^{-9}$ & $1.69 \cdot 10^{-9}$ & $4.29 \cdot 10^{-9}$
  & $1.41 \cdot 10^{-9}$ & $6.64 \cdot 10^{-9}$ & $6.23 \cdot 10^{-9}$
  & $5.95 \cdot 10^{-8}$ & $3.22 \cdot 10^{-10}$
  & $1.01 \cdot 10^{-8}$ & $4.14 \cdot 10^{-6}$
  \\
 $\sigma_p^{SD}$ std [pb] 
  & $4.90 \cdot 10^{-5}$ & $4.31 \cdot 10^{-4}$ & $3.59 \cdot 10^{-4}$
  & $9.16 \cdot 10^{-4}$ & $1.80 \cdot 10^{-6}$ & $2.63 \cdot 10^{-3}$
  & $8.83 \cdot 10^{-4}$ & $5.26 \cdot 10^{-7}$
  & $6.10 \cdot 10^{-6}$ & $3.47 \cdot 10^{-5}$
   \\
 $\sigma_p^{SI}$ dn [pb] 
  & $4.08 \cdot 10^{-9}$ & $1.69 \cdot 10^{-9}$ & $4.29 \cdot 10^{-9}$
  & $1.41 \cdot 10^{-9}$ & $6.75 \cdot 10^{-9}$ & $6.23 \cdot 10^{-9}$
  & $5.92 \cdot 10^{-8}$ & $3.22 \cdot 10^{-10}$
  & $1.01 \cdot 10^{-8}$ & $4.14 \cdot 10^{-6}$
    \\
 $\sigma_p^{SD}$ dn [pb] 
  & $4.90 \cdot 10^{-5}$ & $4.31 \cdot 10^{-4}$ & $3.59 \cdot 10^{-4}$
  & $9.16 \cdot 10^{-4}$ & $1.79 \cdot 10^{-6}$ & $2.63 \cdot 10^{-3}$
  & $8.83 \cdot 10^{-4}$ & $5.26 \cdot 10^{-7}$
  & $6.02 \cdot 10^{-6}$ & $3.40 \cdot 10^{-5}$ \\
$N_{\gamma \mbox{~cont.}}(\sigma v)$ [$10^{-29}$ cm$^3$ s$^{-1}$] 
  & $7.23 \cdot 10^{4}$ & $2.90 \cdot 10^{4}$ & $3.05 \cdot 10^{4}$
  & $1.39 \cdot 10^{4}$ & $4.14 \cdot 10^{4}$ & $3.93 \cdot 10^{2}$
  & $6.20 \cdot 10^{2}$ & $3.45 \cdot 10^{4}$
  & $8.04 \cdot 10^{4}$ & $4.64 \cdot 10^{4}$
 \\
$2(\sigma v)_{\gamma \gamma}$ [$10^{-29}$ cm$^3$ s$^{-1}$] 
& 0.073 & 0.57 & 0.64 & 0.67 & 3.21 & 1.21 & 3.39 & 5.07 & 0.0077 & 0.059\\
$(\sigma v)_{Z \gamma}$ [$10^{-29}$ cm$^3$ s$^{-1}$] 
& 0.44 & 1.95 & 2.21 & 1.49 & 4.53 & 1.22 & 3.36 & 4.91 & 5.41 & 0.11 \\
$\Phi_{e^+}$ [GeV$^{-1}$ cm$^{-2}$ s$^{-1}$ sr$^{-1}$] 
  & $5.65 \cdot 10^{-9}$ & $9.86 \cdot 10^{-7}$ & $8.70 \cdot 10^{-7}$
  & $1.32 \cdot 10^{-6}$ & $2.48 \cdot 10^{-8}$ & $5.97 \cdot 10^{-8}$
  & $8.33 \cdot 10^{-8}$ & $1.91 \cdot 10^{-8}$
  & $4.20 \cdot 10^{-7}$ & $7.78 \cdot 10^{-7}$
  \\
$\Phi_{\bar{p}}$ [GeV$^{-1}$ cm$^{-2}$ s$^{-1}$ sr$^{-1}$] 
  & $1.25 \cdot 10^{-8}$ & $3.95 \cdot 10^{-8}$ & $3.17 \cdot 10^{-8}$
  & $8.16 \cdot 10^{-8}$ & $1.40 \cdot 10^{-10}$ & $3.13 \cdot 10^{-9}$
  & $3.68 \cdot 10^{-9}$ & $8.09 \cdot 10^{-11}$
  & $1.25 \cdot 10^{-8}$ & $2.05 \cdot 10^{-8}$ 
   \\ 
$\Phi_{\bar{D}}$ [GeV$^{-1}$ cm$^{-2}$ s$^{-1}$ sr$^{-1}$] 
  & $4.52 \cdot 10^{-13}$ & $8.15 \cdot 10^{-13}$ & $5.34 \cdot 10^{-13}$
  & $3.85 \cdot 10^{-12}$ & $3.07 \cdot 10^{-15}$ & $1.60 \cdot 10^{-13}$
  & $1.57 \cdot 10^{-13}$ & $2.06 \cdot 10^{-15}$
  & $6.55 \cdot 10^{-13}$ & $4.33 \cdot 10^{-13}$
    \\ \hline
\end{tabular}
  \caption{Same as Table \protect\ref{tab:bat-bench}, but for the MSSM benchmark models of our sample scan.}\label{tab:mssm}
\end{sidewaystable}
    
As seen in the table, the rates are typically slightly higher than for the mSUGRA benchmark models considered above. Please remember though, that this set of MSSM benchmark models is achieved with a rather small scan over the MSSM parameter space. Models with even higher rates would be possible to find with more extensive scans of the parameter space.

Many of our models produce fluxes from the Sun that exceed the future sensitivity of
50--1000 events km$^{-2}$ yr$^{-1}$ and hence would be detectable. For the Earth, on the other hand, the fluxes in this set of models are typically much lower than the projected future limit of 20 events km$^{-2}$ yr$^{-1}$. Most of the selected models are potentially detectable with future direct detection experiments as they have a spin-independent scattering cross section larger than $10^{-9}$ pb. Note the complementarity between the direct detection signal and the neutrino flux from the Sun. For example, for model 9, the spin-independent scattering cross section is very close to the sensitivity of future detectors, whereas the neutrino-flux from the Sun is clearly above expected future sensitivities of neutrino telescopes like Antares or IceCube. The gamma-ray yields are in general larger than for the mSUGRA benchmark points, but the issue regarding detectability is still difficult to address as it is strongly affected by the halo model dependence. The positron and antiproton fluxes are generally enhanced as well,
but still well below measured fluxes. Finally, there are several models among those we selected which have an antideuteron flux in the energy range 0.1--0.4 GeV exceeding
$2.6 \times 10^{-13}$ GeV$^{-1}$ cm$^{-2}$ s$^{-1}$ sr$^{-1}$, the expected sensitivity
of the GAPS detector~\cite{GAPSproposal}. 

\subsection{Discussion}

The potential of \ds\ package has been illustrated here, for a few sample models and in two popular scenarios. The code provides the state of the art calculation of the neutralino relic abundance, and allows for the estimate of several quantities of interest for dark matter searches. It has been shown that the code has a very flexible structure in the definition of the supersymmetric dark matter candidate, as well as a rather broad freedom in the choice of the relevant astrophysical setups. Outputs are in simple and general formats, which, as we have shown, can be very easily compared to current and future sensitivities. Some trends on predictions can be extracted from benchmark models, as we have done for some of the mSUGRA models proposed in \cite{bat-bench}, and with sample models selected according to their relic abundance in a more general low energy scan (with the latter being more promising than the former). One should keep in mind however that firmer statements are possible just in light of dedicated and more extensive scans.

\section{Conclusions}

We have here described the computer package \ds, that can be used to calculate various quantities of interest for supersymmetric dark matter searches. We have gone through great efforts and used state-of-the-art techniques to obtain a package that can deliver very accurate results in a flexible setting. We also believe that this package can be of great use for the physics community. 

In this paper we have described briefly the ingredients of \ds\ and shown, with some examples, what one can calculate with it. We encourage the reader to download \ds\ and start using it.

\section*{Acknowledgements}

We thank all our users of previous releases of \ds\ for their feedback. 
The work of L.B. was partially supported by the Swedish Research Council
(VR). J.E. was supported by the Swedish Research Council. P.U.\ was supported in part by the RTN project under grant HPRN-CT-2000-00152 and by the Italian INFN under the
project ``Fisica Astroparticellare''.

\appendix

\section{Included coannihilations}

\label{sec:processes}
In this Appendix, we tabulate the coannihilation processes computed in \ds.
For more details, see \cite{coann2}.

In the table, we  list all $2 \rightarrow 2$ tree-level coannihilation processes 
with sfermions, charginos and neutralinos.  
All the processes are included in the \ds\ code.

It should be noted that we have not included all flavour-changing 
charged current diagrams. The \ds\ vertex code for the charged current 
couplings is written in a general form that includes all possible 
flavour-changing (and flavour-conserving) vertices. The 
flavour-conserving couplings are much larger than the flavour-changing. 
For the sfermion coannihilations with charged currents we only take the 
flavour-conserving contributions, while for the chargino coannihilations 
we include the flavour-changing contributions as well. In a future 
version of \ds, we may as well include the flavour-changing processes 
for the sfermion coannihilations, even if they are not expected to
be important.

We have used the notation $\ft$ for sfermions and $f$ for
fermions. Whenever the isospin of the sfermion/fermion is important,
it is indicated by an index $u$ ($T_3 = 1/2$) or $d$ ($T_3 = -1/2$).
The sfermions have an additional mass eigenstate index, that can take
the values 1 and 2 (except for the sneutrinos which only have one mass
eigenstate).  A further complication to the notation is when the
sfermions and fermions in initial, final and exchange state can belong
to different families. Primes will be used to indicate when we have
this freedom to choose the flavour. So, e.g.~$\ft_u$ and $f_u$ will
belong to the same family while $\ft_u$ and ${f}^{\prime}_u$ can belong to the
same or to different families. Note that the colour index of (s)quarks
as well as gluons ($g$) and gluinos ($\tilde{g}$) is suppressed.

Besides the sfermions we also have neutralinos and charginos in the
initial states. The notation used for these are the following. The
neutralinos are denoted by $\chi^0_j$ with the index running from 1 to
4.  The charginos are similarly denoted $\chi^\pm_j$ with the index
taking the values 1 and 2.

In the table,  a common notation is introduced for
gauge and Higgs bosons in the final state. We denote these with $B$
with an upper index indicating the electric charge. So $B^0$ means
$H^0_1, H^0_2, H^0_3, Z, \gamma$ and $g$ while $B^\pm$ is $H^\pm$ and
$W^\pm$.  We will use additional lower indices $m$ and $n$ when we
have more than one boson in the final state. Thus indicating that the
bosons can be either different or identical. Note that the case of two
different bosons also includes final states with one gauge boson and
one Higgs boson.

The table has been made very general. This means that when a set of
initial and final state (s)particles has been specified, the given
process might not run through all the exchange channels listed for the
generic process. Exceptions occur whenever an exchange (s)particle
does not couple to the specific choice of initial and/or final
state. As an example we see that since the photon does not couple to
neutral (s)particles, none of the exchange channels listed for the
generic process $\ft_i + \chi^0_j \into B^0 + f$ actually exist for
the specific process $\snu + \chi^0 \into \gamma + \nu$. All these
exceptions can be found in the extended tables in Ref.\
\cite{ds-manual}. Also note that the list of processes is not complete
with respect to trivial charge conjugation. For each process of
non-vanishing total electric charge in the initial state there exists 
another process which is obtained by charge conjugation.

\begin{table}
\begin{tabular}{p{4cm}p{3cm}p{2cm}p{2.0cm}p{0.5cm}}
 & \multicolumn{4}{c}{Diagrams} \\ \cline{2-5}
Process & s  & t & u & p \\
\hline
~ & \\[-2.5ex]
$\chi^0_i \chi^0_j \into B^0_m B^0_n$ & $H^0_{1,2,3},Z$ & $\chi^0_k$ & $\chi^0_l$ \\
$\chi^0_i \chi^0_j \into B^-_m B^+_n$ &Ê$H^0_{1,2,3},Z$ & $\chi^+_k$ & $\chi^+_l$ \\
$\chi^0_i \chi^0_j \into f \bar{f}$ &Ê$H^0_{1,2,3},Z$ & $\ft_{1,2}$ & $\ft_{1,2}$ \\[1ex]
\hline
~ & \\[-2.5ex]
$\chi^+_i \chi^0_j \into B^+_m B^0_n$ & $H^+,W^+$ & $\chi^0_k$ & $\chi^+_l$ \\
$\chi^+_i \chi^0_j \into f_u \bar{f}_d$ &Ê$H^+,W^+$ & $\ft^\prime_{d_{1,2}}$ & $\ft^\prime_{u_{1,2}}$ \\[1ex] \hline
~ & \\[-2.5ex]
$\chi^+_i \chi^-_j \into B^0_m B^0_n$ & $H^0_{1,2,3},Z$ & $\chi^+_k$ & $\chi^+_l$ \\
$\chi^+_i \chi^-_j \into B^+_m B^-_n$ & $H^0_{1,2,3},Z,\gamma$ & $\chi^0_k$ & {} \\
$\chi^+_i \chi^-_j \into f_u \bar{f}_u$ &Ê$H^0_{1,2,3},Z,\gamma$ & $\ft^\prime_{d_{1,2}}$ & {} \\
$\chi^+_i \chi^-_j \into \bar{f}_d f_d$ &Ê$H^0_{1,2,3},Z,\gamma$ & $\ft^\prime_{u_{1,2}}$ & {} \\
$\chi^+_i \chi^+_j \into B^+_m B^+_n$ & {} & $\chi^0_k$ & $\chi^0_l$ \\[1ex]
\hline
~ & \\[-2.5ex]
$\ft_i \chi^0_j     \into B^0 f$   & $f$   & $\ft_{1,2}$ & $\chi^0_l$ \\
$\ft_{d_i} \chi^0_j \into B^- f_u$ & $f_d$ & $\ft_{u_{1,2}}$ & $\chi^+_l$    \\
$\ft_{u_i} \chi^0_j \into B^+ f_d$ & $f_u$ & $\ft_{d_{1,2}}$ & $\chi^+_l$ \\[1ex]
\hline
~ & \\[-2.5ex]
$\ft_{d_i} \chi^+_j \into B^0 f_u$ & $f_u$ & $\ft_{d_{1,2}}$ & $\chi^+_l$    \\
$\ft_{u_i} \chi^+_j \into B^+ f_u$ & {}    & $\ft_{d_{1,2}}$ & $\chi^0_l$    \\
$\ft_{d_i} \chi^+_j \into B^+ f_d$ & $f_u$ & {}              & $\chi^0_l$    \\
$\ft_{u_i} \chi^-_j \into B^0 f_d$ & $f_d$ & $\ft_{u_{1,2}}$ & $\chi^+_l$ \\
$\ft_{u_i} \chi^-_j \into B^- f_u$ & $f_d$ & {}              & $\chi^0_l$ \\
$\ft_{d_i} \chi^-_j \into B^- f_d$ &       & $\ft_{u_{1,2}}$ & $\chi^0_l$ \\[1ex]
\hline
~ & \\[-2.5ex]
$\ft_{d_i} \ft_{d_j}^\ast \into B^0_m B^0_n$ & $H^0_{1,2,3},Z,g$ & $\ft_{d_{1,2}}$    &
$\ft_{d_{1,2}}$ & p \\
$\ft_{d_i} \ft_{d_j}^\ast \into B^-_m B^+_n$ & $H^0_{1,2,3},Z,\gamma$ & $\ft_{u_{1,2}}$ &
{}              & p \\
$\ft_{d_i}\ft_{d_j}^{\prime\ast}\into f^{\prime\prime}_d \bar{f}^{\prime\prime\prime}_d$ & $H^0_{1,2,3},Z,\gamma,g$ & $\chi^0_k,\gt$ &
{}              & {}  \\
$\ft_{d_i} \ft_{d_j}^{\prime\ast} \into f^{\prime\prime}_u \bar{f}^{\prime\prime\prime}_u$ & 
$H^0_{1,2,3},Z,\gamma,g$ & $\chi^+_k$ & 
{}              & {}  \\
$\ft_{d_i}{\ft}^\prime_{d_j} \into f_d {f}^\prime_d$ & {} & $\chi^0_k,\gt$  & 
$\chi^0_l,\gt$& {} \\[1ex]
\hline
~ & \\[-2.5ex]
$\ft_{u_i}\ft_{d_j}^\ast\into B^+_mB^0_n$& $H^+,W^+$ & $\ft_{d_{1,2}}$ & $\ft_{u_{1,2}}$ & p \\
$\ft_{u_i}\ft_{d_j}^{\prime\ast}\into f^{\prime\prime}_u \bar{f}^{\prime\prime\prime}_d$ & $H^+,W^+$ & $\chi^0_k,\gt$ & {} & {}\\
$\ft_{u_i}{\ft'}_{d_j}\into f^{\prime\prime}_u f^{\prime\prime\prime}_d$ & {} & $\chi^0_k,\gt$ & $\chi^+_l$ & {} \\[1ex]
\hline
\end{tabular}
\caption{Included coannihilation processes through $s-$, $t-$, $u-$channels 
and four-point interactions (p). For the $\ft_{d_i} \ft_{d_j}^{(\ast)}$ processes
the corresponding process for up-type sfermions can be obtained by interchanging
the $u$ and $d$ indices.}
\label{tab:coanns}
\end{table}
\newpage


\end{document}